\documentclass[twocolumn]{aastex63}

\received{----}
\revised{----}
\accepted{----}
\submitjournal{ApJ}

\shorttitle{Dust Reverberation Mapping of 3C\,273}
\shortauthors{Sobrino Figaredo et al.}

\graphicspath{{./}{revised_figures/}}

\begin{document}


\title{Dust Reverberation of 3C\,273: torus structure and lag - luminosity relation}


\author{Catalina Sobrino Figaredo}
\affiliation{Astronomisches Institut Ruhr-Universit\"at Bochum, Universit\"atsstr. 150, D-44801 Bochum, Germany}

\author{Martin Haas}
\affiliation{Astronomisches Institut Ruhr-Universit\"at Bochum, Universit\"atsstr. 150, D-44801 Bochum, Germany}

\author{Michael Ramolla}
\affiliation{Astronomisches Institut Ruhr-Universit\"at Bochum, Universit\"atsstr. 150, D-44801 Bochum, Germany}

\author{Rolf Chini}
\affiliation{Astronomisches Institut Ruhr-Universit\"at Bochum, Universit\"atsstr. 150, D-44801 Bochum, Germany}
\affiliation{Universidad Cat\'{o}lica del Norte, Antofagasta, Chile}

\author{Julia Blex}
\affiliation{Astronomisches Institut Ruhr-Universit\"at Bochum, Universit\"atsstr. 150, D-44801 Bochum, Germany}

\author{Klaus Werner Hodapp}
\affiliation{Institute for Astronomy, 640 North A'oh\={o}k\={u} Place, Hilo, HI 96720-2700, USA}

\author{Miguel Murphy}
\affiliation{Universidad Cat\'{o}lica del Norte, Antofagasta, Chile}

\author{Wolfram Kollatschny}
\affiliation{Institut f\"ur Astrophysik, Universit\"at G\"ottingen, Friedrich-Hund Platz 1, D-37077 G\"ottingen, Germany}

\author{Doron Chelouche}
\affiliation{Physics Department and the Haifa Research Center for Theoretical Physics and Astrophysics, University of Haifa, Israel}

\author{Shai Kaspi}
\affiliation{School of Physics \& Astronomy and the Wise Observatory, The Raymond and Beverly Sackler Faculty of Exact Sciences Tel-Aviv University, Israel}

\begin{abstract}

 We monitored the $z =$ 0.158 quasar 3C\,273 between 2015 and 2019 in the optical ($BVrz$) and near-infrared (NIR, $JHK$) with the aim to perform dust reverberation mapping. Accounting for host galaxy and accretion disk contributions, we obtained pure dust light curves in $JHK$. 
Cross correlations between the $V$-band and the dust light curves yield an average rest-frame delay for the hot dust of \hbox{$\tau_{\rm cent}~\sim$ 410 days}. This is a factor 2 shorter than expected from the the dust ring radius \hbox{$R_{\rm x}\sim$ 900~light days reported from interferometric studies.} The dust covering factor (CF) is about 8\%, much smaller than predicted from the half covering angle of $45^{\circ}$ found for active galactic nuclei (AGN). We analyse the asymmetric shape of the correlation functions and explore whether an inclined bi-conical bowl-shaped dust torus geometry could bring these findings ($\tau_{\rm cent}$, $R_{\rm x}$ and CF) into a consistent picture. The hot varying dust emission originates from the edge of the bowl rim with a small covering angle $40^{\circ} < \theta < 45^{\circ}$, and we see only the near side of the bi-conus. Such a dust gloriole with $R_{\rm x} = 900\pm200$~ld and an inclination $12^{\circ}$ matches the data remarkably well. Comparing the results of 3C\,273 with literature for less luminous AGN, we find a lag--luminosity relation $\tau \propto L^{\alpha}$ with $\alpha = 0.33 - 0.40$, flatter than the widely adopted relation with \hbox{$\alpha \sim$ 0.5.} We address several explanations for the 
new lag--luminosity relation.

\end{abstract}





\section{Introduction}
\label{sec:introduction}

The quasar paradigm comprises a supermassive black hole
(SMBH), a central X-ray source, an accretion disk (AD),
surrounded by a broad line region (BLR), and a molecular dusty
torus (TOR) farther out. The three components
AD, BLR and TOR may have smooth transitions between each other rather
than being separated entities with sharp boundaries.
Of particular interest here is the 3-dimensional geometry of the
central region and the three components.

As the inner quasar regions cannot be resolved by conventional
imaging techniques, reverberation mapping (RM) is the main tool of
the trade \citep{1972ApJ...171..467B,1973ApL....13..165C, 1986ApJ...305..175G,1993PASP..105..247P,2004PASP..116..465H}.
RM traces the delayed response of irradiated regions to the light
fluctuations of the continuum emission from the inner AD. 
As a first approximation, the size
of the irradiated region can be inferred
from the time lag $\tau$.
This way, near-infrared (NIR) RM studies of the dusty torus find a radius
$R_{\tau} = c \cdot \tau$
\citep{1989ApJ...337..236C,2006ApJ...639...46S}.

\subsection{The size - luminosity relation}
\label{sec:rl_relation}

A remarkable finding is the relation between the reverberation based size, $R_{\tau}$, and the AGN luminosity, $L$, with
$R_{\tau} \propto L^\alpha$ and $\alpha \approx 0.5$
\citep{2006ApJ...639...46S,2007arXiv0711.1025G,2014ApJ...788..159K,2019arXiv191008722M}.
Such a relation with $\alpha \approx 0.5$ has been expected, if the hot NIR emitting dust  is located in a
simple equatorial geometry and $R_{\tau}$ measures the dust
sublimation radius, $R_{\rm sub}$, inferred from the UV luminosity \citep{Barvainis87}.
This relation may hold the key for cosmological applications to measure quasar distances
and to check for dark energy
\citep{1998SPIE.3352..120K,1999OAP....12...99O,2002ntto.conf..235Y,2014ApJ...784L..11Y,2014ApJ...784L...4H}.
However, $R_{\tau}$ is about 3 times smaller than $R_{\rm sub}$
\citep{2007A&A...476..713K,2014ApJ...788..159K}.
This suggests that internal effects like from the 3-dimensional dust geometry
or from a reduced heating of potentially shielded dust clouds \citep{2007arXiv0711.1025G}
need to be better understood, before cosmological implications from the $R-L$ relation should be drawn.

Before we address such effects, we note that a similar $R-L$
relation had been found between the size of the broad line region (BLR) and the AGN luminosity: in the first observational estimate of the R-L relationship \cite{1991ApJ...370L..61K} found a slope $\alpha = 0.33 \pm 0.2$ for C\,IV. 
Subsequently, for $H\beta$ \cite{2000ApJ...533..631K} report  $\alpha = 0.7 \pm 0.03$ and then \cite{2013ApJ...767..149B} after correction of the host galaxy contribution refined this slope to $\alpha = 0.533 \pm 0.034$.
One more piece to the puzzle was added by 
\cite{2016ApJ...825..126D} reporting  for AGN with high accretion rates a shallow slope $\alpha \approx 0.3$.
Possible explanations for that have been considered,
involving the ionisation parameter \citep{2019ApJ...870...84C}.
Notably, nearly half a century ago 
\cite{1977SvAL....3....1D} used the Str\"omgren-type argument and a value of $\alpha = 0.33$ 
for calculating the first single-epoch AGN black hole masses, which agree remarkably well with the latest reverberation based estimates
\citep{2009AstL...35..287B}.

\subsection{The torus structure}
\label{sec:torus_structure}

Interferometric $K-$band measurements with the KECK interferometer
\citep{2009A&A...507L..57K,2011A&A...527A.121K} and
the VLTI/Gravity instrument \citep{2019arXiv191000593G} resolved the innermost
dusty structure of 15 AGN (8 with KECK, 8 with VLTI, and 3C\,273 in common).
The effective ring radii $R_{ \rm ring}$ derived from the observed visibilities
scale roughly with the AGN luminosity $L^{1/2}$, but the \cite{2019arXiv191000593G}
report a relative size decline in the two highest luminosity AGN.
For AGN with available $K-$band reverberation measurements,
$R_{ \rm ring}$ is, on average, larger than $R_{\tau}$.
\cite{2011A&A...527A.121K} suggested  that the interferometric measurements at least partly
resolve the dust sublimation zone,
and that the ratio \hbox{$r = R_{ \rm ring} / R_{\tau}$} yields information on
the compactness ($r \approx 1$) or extent ($r > 1$) of the hot dust distribution.
For the entire sample, however, $R_{ \rm ring}$ is systematically smaller than $R_{\rm sub}$.
To match  $R_{\rm ring}$ with $R_{\rm  sub}$, \cite{2007A&A...476..713K,2009A&A...507L..57K}
suggested that the dust grain sizes are larger or that the central engine
radiation is significantly anisotropic.

\cite{2010ApJ...724L.183K,2011ApJ...737..105K} proposed a bowl-shaped dust torus which smoothly continues into the central AD. In their model, the AD emission is anisotropic, being
highest toward the polar region and lowest toward the
equatorial region. The anisotropy of the AD emission controls the angle-dependent dust sublimation
radius and thus the concave rim of the bowl,
allowing for parts of the dust to lie closer to the AD than $R_{\rm sub}$
calculated from the luminosity towards the polar direction.
Despite attractive features, however, this model needs to be expanded to account for
the influence of the BLR onto the entire system \citep{2012MNRAS.426.3086G}.

For the well studied Seyfert-1 galaxy NGC\,5548, \cite{2007arXiv0711.1025G}
have analyzed the AGN energy budget and derived crucial
constraints on the geometry of the BLR and the dust torus:
the BLR has a likely covering factor about 40\%, which translates
to a half covering angle $\theta \approx 25^\circ$, as measured
from the equatorial plane.
The BLR shields a substantial
fraction of the dust torus from direct illumination by the AD,
allowing for the observed relatively small ($<$20\%) NIR
contribution to the AGN energy budget.

\cite{2012MNRAS.426.3086G} combined the findings by
\cite{2007arXiv0711.1025G} and \cite{2010ApJ...724L.183K,2011ApJ...737..105K}
into a BLR-TOR system
confined by a paraboloidal bowl-shaped torus rim where
the BLR clouds lie close to and above the rim (their Fig. 1).
The observer sees only the emission from the near side of the bi-polar bowl.
The BLR clouds
shield part of the dust from the AD radiation at $\theta \lesssim 40^\circ$, so that
the hot dust emission arises essentially from  the top rim of
the bowl at $40^\circ < \theta < 45^\circ$.
\cite{2012MNRAS.426.3086G} tested their model
with reverberation data of NGC 5548.

Based on high-cadence dust RM observations of the Sy-1 WPVS\,48, \cite{2014A&A...561L...8P}
found an exceptionally sharp NIR echo, which led
them to favour that the varying hot dust is essentially located
at the edge ($40^\circ \lesssim \theta \lesssim 45^\circ$)
rather than along the entire bowl rim which crosses a large range of iso-delay surfaces.
In a slightly generalized view \cite{2015OAP....28..175O} proposed that the hot dust emission
comes from the near side of a hollow bi-conical outflow
which is co-spatial with the iso-delay paraboloids, a case requiring that the
inclination to the line-of-sight is small.
Based on observations of the Sy-1 galaxy 3C\,120, with a jet inclination $i\,\sim\,16^\circ$,
\cite{2018A&A...620A.137R} found that the hot dust echo is relatively sharp
and symmetric in contrast to the more complex broad H$\alpha$ echo.
This is consistent with a paraboloidal bowl model where the BLR is spread over
many iso-delay surfaces,
yielding the smeared and structured  H$\alpha$ echo as observed,
while the hot dusty bowl-edge matches a relatively narrow range of iso-delay contours.
The important feature of such a gloriole-like dust emission is the geometric foreshortening effect of the reverberation signal,
because the dust lies above the equatorial plane and closer to the observer.
While such structure studies have been performed for Seyfert galaxies, i.e. low luminosity AGN,
they should be extended to higher luminosity quasars, in order to recognize luminosity-dependent trends.

\subsection{3C\,273}
\label{sec:this_target}

3C\,273 at $z\,=\,0.158$ is the most luminous nearby quasar and may serve as a bench mark for any luminosity-dependent relations.
3C\,273 is the brightest quasar making it an ideal target for observations of any kind.

Based on their 7 years long RM campaign \cite{2000ApJ...533..631K}
determined rest frame Balmer line lags of
$\tau (H\alpha) = 443$\,d, $\tau (H\beta) = 330$\,d and $\tau (H\gamma) = 265$\,d versus the $B-$band; here $\tau$ is the centroid of the interpolated cross correlation function. 
\cite{2011A&A...527A.121K} and \cite{2019arXiv191000593G} reported a dust ring radius of $R \sim 900$\,ld inferred from interferometric $K-$band measurements.

Yet, to our knowledge, no dedicated dust reverberation mapping campaign of 3C\,273 has been performed.
Fortunately, \cite{2008A&A...486..411S} have collected all published photometry of 3C\,273 of
the past 50 years in the ISDC\footnote{http://isdc.unige.ch/3c273/}, Geneva,
adjusted the photometry from different telescopes, removed flares, selected the best data.
Cross correlating the $JHK$ with optical and UV light curves, they derived
a rest frame dust lag of $\tau (K) \sim 1 \pm 0.2$ years.
An important part of these data are the light curves obtained by
\cite{1999AJ....118...35N}  at the 5\,m Palomar Hale telescope between 1974 and 1998,
using a single-element InSb photovoltaic detector and
chopping/nodding technique to remove the effects of the sky emission;
as described by Soldi et al. these data unfortunately were not available in tabulated form
but extracted from the figure.

Remarkably the dust lag (365\,d) is shorter than the H$\alpha$ lag (443\,d),
a fact which at first glance suggests the dust torus to be smaller than the BLR, hence
appears in contradiction to the AGN unified scheme \citep{1993ARA&A..31..473A}.
This led us to embark on a dedicated reverberation campaign of 3C\,273,
and we here report on the results of the dust reverberation, a re-analysis of Soldi et al.'s data, a comparison with bowl-shaped models and the interferometric measurements, and the impact on the lag--luminosity relation.

Section \ref{sec:observations} describes the observations and data.
Section \ref{sec:results} shows the light curves, the determination and subtraction
of the host galaxy and the contribution of the AD to the NIR light curves,
the properties of the varying dust emission,
the dust time lag and the derivation of the dust covering factor.
Section \ref{sec:bowl_shaped} considers the RM results in the framework of a bowl-shaped TOR
geometry using the additional constraints of the interferometric size and the small
dust covering factor.
Section \ref{sec:discussion} discusses the findings and implications.
Section \ref{sec:conclusions} provides a summary and conclusions.
Throughout this paper we adopt $\Omega_{\rm m} = 0.27$, $\Omega_{\rm v}= 0.73$
and H$_\circ$\,=\,73\,km\,s$^{-1}$\,Mpc$^{-1}$
yielding an angular-size distance of 546\,Mpc (luminosity distance of 734\,Mpc) for 3C\,273.

\section{Observations and data reduction}
\label{sec:observations}

3C\,273 was observed between April 2015 and June 2019 with semi-robotic optical and
NIR telescopes of the Bochum University Observatory near Cerro Armazones (OCA)in Chile\footnote{https://en.wikipedia.org/wiki/Cerro\_Armazones\_Observatory} \citep{2016SPIE.9911E..2MR}.
The optical telescopes used are: the 40\,cm Bochum Monitoring
Telescope, BMT \citep{2013AN....334.1115R}; the 15\,cm Robotic Bochum Twin Telescope ROBOTT
(formerly named VYSOS\,6, for short V6) \citep{2012AN....333..706H};
the 25\,cm BEST-II \citep{2009A&A...506..569K} and
the 80\,cm infrared telescope IRIS \citep{2010SPIE.7735E..1AH}.

Typically $10 - 20$ dithered images were taken per filter and night and later combined.
All images were reduced by the corresponding instrumental data reduction pipelines
(dark, bias and flat correction).
Astrometric matching was performed with Scamp \citep{Bertin_2006}.
Before stacking multiple exposures, they were resampled onto a common coordinate grid
with $0\farcs75$ pixel size using Swarp \citep{Bertin_2002};
the seeing has typically a point spread function of $\sim$ 3$\arcsec$ FWHM.
The photometry is performed on combined frames with a fixed 7$\farcs$5 aperture
found to be the optimum in our previous studies,
e.g. \citep{2011A&A...535A..73H, 2014A&A...561L...8P, 2018A&A...620A.137R}.

\begin{table}
  \centering
  \caption{Parameters of the 5 years monitoring campaign of 3C\,273.
    Filters, effective wavelengths $\lambda_{\rm eff}$, zero mag flux $f_0$,
    average flux and number of light curve data points (observed nights).
    \label{tab:observations}}
  \begin{tabular}{c c c c c}
    \tableline 
    Filter & $\lambda_{\rm eff}$\,[$\mu$m] & $f_0$\,[Jy] & avg.Flux [mJy] & Obs.nights  \\ \tableline 
    $B$ & 0.433 & 4266.7  &  23.04\,$\pm$\,1.39 & 109\\[1pt]
    $V$ & 0.550 & 3836.3  &  24.34\,$\pm$\,1.12 & 119\\[1pt]
    $r$ & 0.623 & 3631.0  &  23.42\,$\pm$\,1.41 & 128\\[1pt]
    $z$ & 0.906 & 3631.0  &  24.59\,$\pm$\,1.57 & 78\\[1pt]
    $J$ & 1.24  & 1594.0  &  30.09\,$\pm$\,0.92 & 44\\[1pt]
    $H$ & 1.65  & 1024.0  &  41.38\,$\pm$\,1.99 & 24\\[1pt]
    $K$ & 2.16  & 666.7   &  76.71\,$\pm$\,3.42 & 77\\[1pt] \tableline
  \end{tabular}
\end{table}

The light curves were created using $5 - 10$ nearby ($<\,30\arcmin$)
calibration stars. The optical and NIR absolute flux calibration was performed by comparison
with the PANSTARSS and the 2MASS catalog respectively, also airmass dependent
extinction \citep{2011A&A...527A..91P} and Galactic foreground
extinction \citep{2011ApJ...737..103S} corrections were applied.
A summary on the filters (their effective wavelength and zero flux),
the average flux of 3C\,273 and total number of observing nights is
listed in Table~\ref{tab:observations}.

The time sampling of our light curves is quite coarse,
typically with median about $4 - 7$\,d but only over a few months per year.
Since 3C\,273 is bright and is nearly daily monitored in the $B,V$ bands
by AAVSO\footnote{https://www.aavso.org/}
we planned to use these light curves if necessary.
Fortunately, \cite{2019ApJ...876...49Z} kindly sent us their more
comprehensive optical light curves (median $\sim$1 day) between 2008 and 2018,
and we used these light curves in addition to ours for the scientific analysis,
e.g. cross correlations and modeling.


\begin{figure*}
  \centering
    \includegraphics[width=18cm]{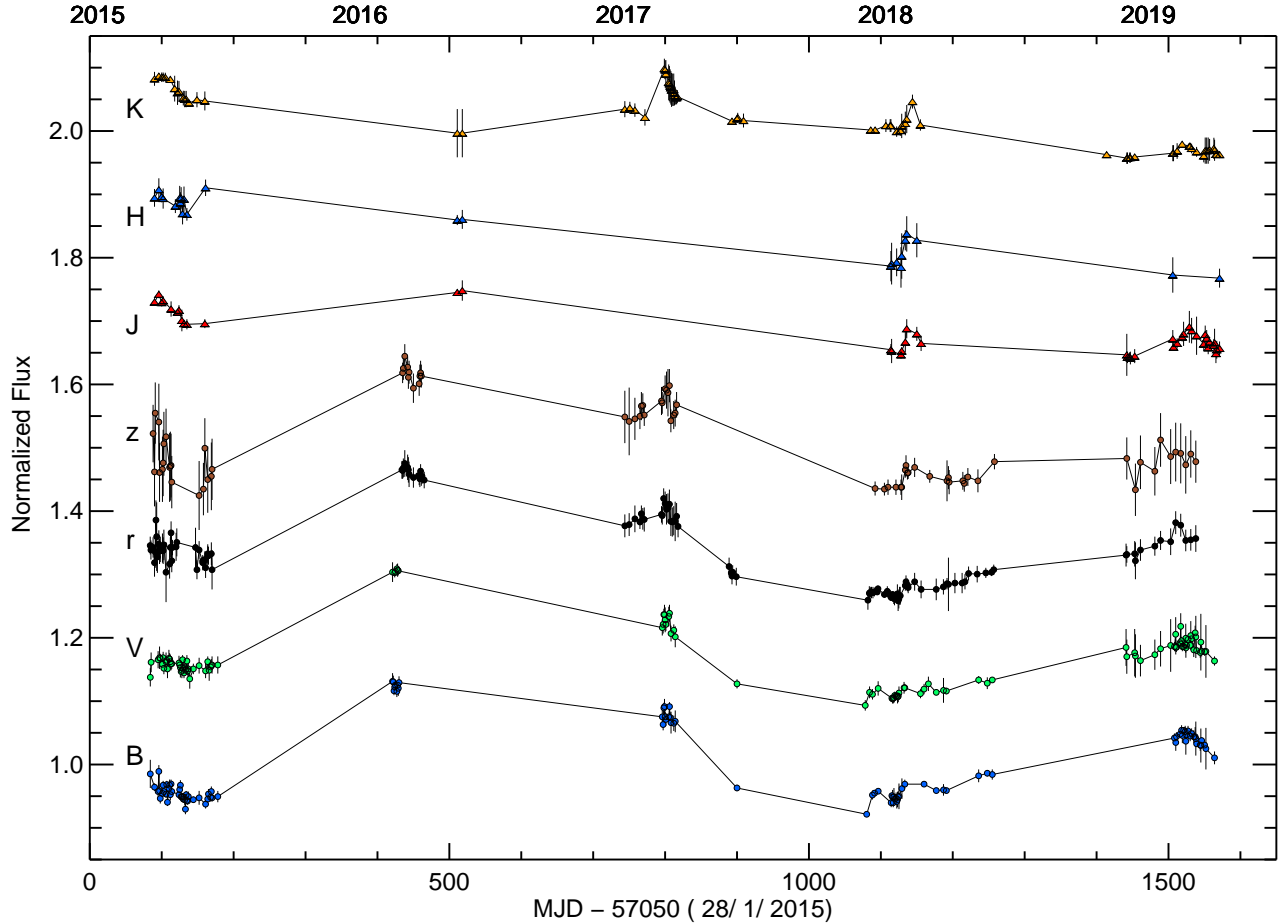}
  \caption{3C\,273 normalized light curves: $BVrz$ represented as circles
and $JHK$ as triangles. All optical filters show the same pronounced
variation features: a 20\% flux increase between 2015 and 2016, followed
by a softer flux decrease of almost 20\% until begin of 2018,
and again an increase of 10\% towards 2019.
For the NIR, note the decrease between 2015 and 2016 in $H$ and $K$ in contrast
to the increase in $J$, suggesting that at least in $J$ the light curve
is strongly contaminated by the accretion disk, while in $K$ the hot dust
emission dominates; $H$ looks like a mixture of AD and dust emission.}
  \label{fig:optical_lc_all}
  \vspace{3mm}
\end{figure*}

\section{Results}\label{sec:results}

\subsection{Optical and near-infrared light curves}

Figure~\ref{fig:optical_lc_all} shows the optical and NIR light curves of 3C\,273.
In all optical filters $BVrz$ (open circles) the variations show the same pronounced features:
between the years 2015 and 2016 the flux
increases by about 20\%, then it decreases by $20 - 25$\% during two years until begin of 2018
where it starts to increase again by about 10\% towards mid of 2019.
For the NIR light curves of 3C\,273 (open triangles): the variations in $J$ resemble
those in the optical light curves but in $H$ and $K$ they differ.
The flux in $J$ increases between 2015 and 2016 but not as pronounced
as in the optical bands, and then it decreases until 2019. For $H$ and $K$ the
trend is different: instead of a flux increase between 2015 and 2016, a decrease is observed.
Then the $K-$band shows an increase towards 2017 and a decrease thereafter.
Unfortunately there was no useful $J$ and $H$ data collected in 2017.
The difference between $JHK$ suggests that at least in $J$ the light curve is
strongly contaminated by the accretion disk, while in $K$ the hot dust emission may dominate;
$H$ looks like an equal mixture of AD and dust emission.
To obtain the pure $JHK$ dust light curves, the contribution of
host galaxy and AD to the NIR bands has to be removed (Sect.~\ref{sec:pure_dust_light_curves}).

\begin{figure}
  \centering
  \includegraphics[width=\columnwidth]{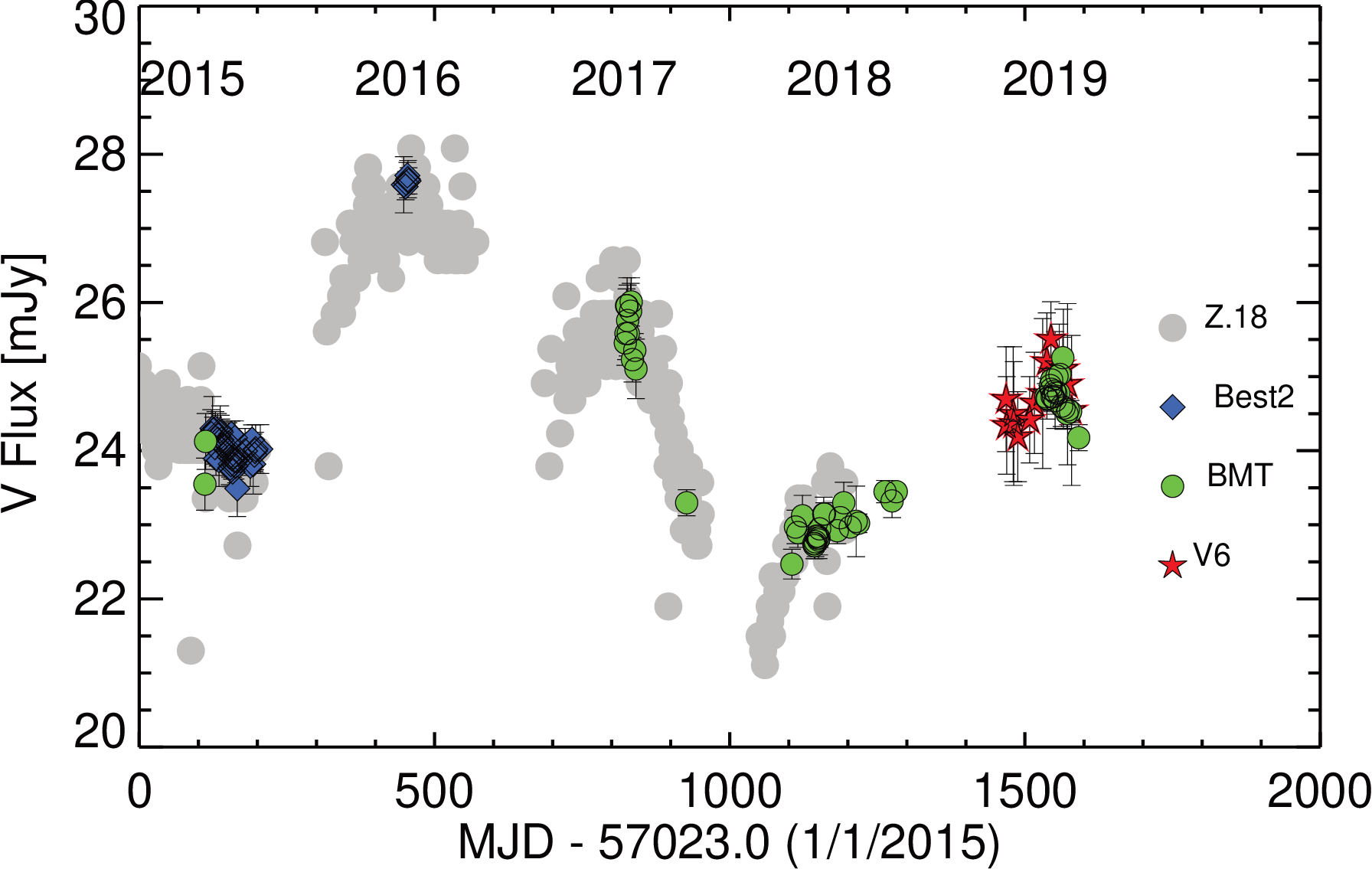}
  \caption{$V-$band light curve plotted with different symbols for the
    three optical telescopes (BMT, BEST-II, ROBOTT=V6).
    The data match with each other and
    with the more comprehensive light curve obtained until March 2018
    by \cite{2019ApJ...876...49Z}, plotted with grey dots.
    \label{fig:optical_lc}
  }
  \vspace{3mm}
\end{figure}

Depending on the telescope availabilities,
the light curves of some filters were obtained  with different telescopes
(BMT, BEST-II, ROBOTT). We checked for telescope-dependent differences between the
light curves.
We found that any differences are smaller than $1 - 2$\%,
and that they are due to an additive offset which increases with the native camera pixel size
(0$\farcs$8 for BMT, 1$\farcs$5 for BEST-II, 2$\farcs$4 for ROBOTT).
This dependence likely comes from a larger host galaxy contribution when the camera has larger
native pixels, despite the resampling to a common pixel size of 0$\farcs$75 
(see Sect.~\ref{sec:observations}).
We corrected for the flux offsets between BMT, BEST-II, ROBOTT and scaled the flux
to that of the BMT;
we note that the inter-telescope corrections were small so that the results,
e.g. on the variability features, are essentially unchanged.
Figure~\ref{fig:optical_lc} shows the $V-$band light curve (after offset correction)
plotted with different symbols for the three optical telescopes.
In addition, grey circles show the $V-$band light curve from the 10 years monitoring
campaign by
\cite{2019ApJ...876...49Z}. This light curve was essentially obtained with $1.5 - 2.5$\,m
class telescopes.
Both ours and Zhang's light curves match excellently within the scatter;
the scatter at a given short time  interval ($\sim$100~d) likely marks the true
photometric light curve
uncertainty. It is similar ($\sim$1\%) for both light curves.

Our light curves are made available in a Journal On-line Table, having five columns:
(1) Filter, (2) Telescope, (3) MJD, (4) Flux [mJy], and (5) Flux error [mJy].

\subsection{Construction of the pure dust light curves}
\label{sec:pure_dust_light_curves}

To construct the pure dust light curves, we determined the host galaxy brightness
in the optical and extrapolated it to the NIR via model SEDs.
Likewise, we determined the AD brightness in the optical and  
extrapolated it to the NIR via a power-law \citep{2005MNRAS.364..640K,2008Natur.454..492K}. 
The pure dust light curves are then obtained from the observed NIR
light curves after subtraction of the NIR host and AD contributions.

\subsubsection{Host galaxy}
\label{sec:host}

Based on $HST$ imaging \cite{1997ApJ...479..642B} found that
3C\,273 has an elliptical host galaxy with morphology type E4
and
from off-nucleus spectroscopy
\cite{2010MNRAS.408..713W} found a contribution of about 14\% young stellar
population and derived a rest frame host galaxy color $B-V = 0.77$.

\begin{figure}
  \centering
  \includegraphics[width=\columnwidth]{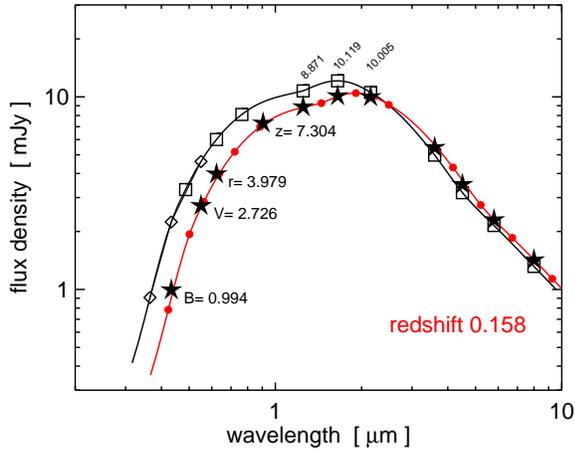}
  \caption{ Construction of the 3C\,273 host SED, based on colors for an elliptical galaxy
    with 14\% flux contribution from a young stellar population: open diamonds
    \citep{1995PASP..107..945F, 2010MNRAS.408..713W}, open squares
    \cite{2006MNRAS.372..199C} and \cite{Jarrett2019}.
    In the rest frame, the black solid line
    depicts a spline function fitted to the open symbols. The red line shows the spline
    shifted to the redshift of 3C\,273 whereby the red dots correspond to the open symbols
    in the rest frame spline.
    The flux scaling of the template is described in the text  (Sect.~\ref{sec:host}).
    The black filled stars on the red line mark the predicted observed fluxes in
    the filters of interest with values as labelled.
    \label{fig:host_sed}
  }
  \vspace{3mm}
\end{figure}


With these constraints at hand, we constructed a restframe host SED template for 3C\,273 based
on the colors of an elliptical host galaxy, as determined for $UBV$ by
\cite{1995PASP..107..945F}, for Sloan $gri$ filters and $JK$ filters by
\cite{2006MNRAS.372..199C}, 
and for 2MASS $JHK_{\rm s}$ and $Spitzer/IRAC$ filters by \cite{Jarrett2019}.
Because of the presence of young stars \citep{2010MNRAS.408..713W}, we used slightly bluer
colors $UBV$ and $r-i$ and $r-J$ and a slightly shallower 1.6\,$\mu$m bump.
Figure~\ref{fig:host_sed} shows the resulting SED template.
It is then fit by a spline function (black solid line) and the spline function
is shifted to the redshift $z=0.158$ of 3C\,273 (red solid line).
Then we sampled the redshifted spline function at the observed wavelengths of interest
(filled black star symbols) and derived the host flux
ratios $B/V$ and $r/z$ in the observer's frame for use in the Flux Variation Gradient (FVG) analysis.

\begin{figure}
  \centering
  \includegraphics[width=0.85\columnwidth]{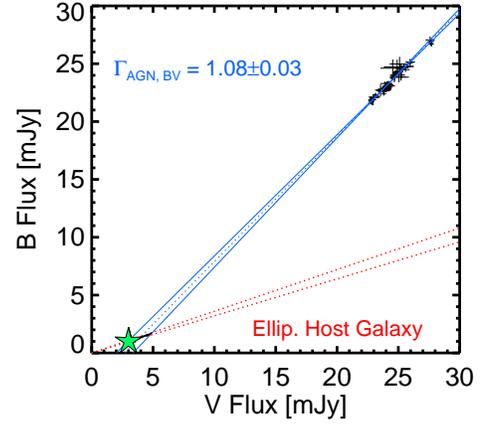}
  \includegraphics[width=0.85\columnwidth]{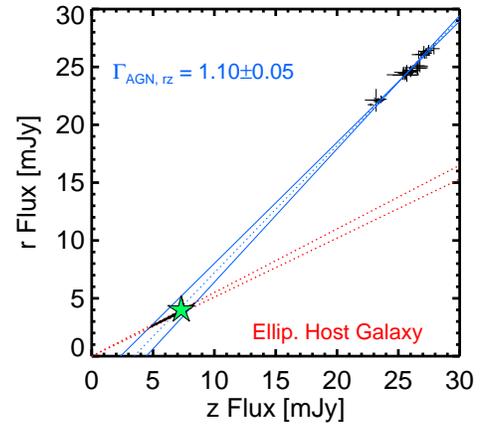}
  \caption{$B/V$ and $r/z$ flux-flux diagrams. 
    Black crosses indicate the matched fluxes for every night with their errors, 
    blue lines the AGN slope $\pm$ error and the red dotted lines mark the host flux ratios
    for an elliptical galaxy derived from the SED in Figure~\ref{fig:host_sed}.
    The derived $BVrz$ host fluxes are plotted as a green star.
    \label{fig:fvg_bv}
  }
  \vspace{3mm}
\end{figure}

To estimate the host contribution in our data, we applied the FVG  method
proposed by \cite{1981AcA....31..293C}, further established by \cite{1992MNRAS.257..659W} and
\cite{2010ApJ...711..461S}, and successfully applied by, e.g.,  \cite{2011A&A...535A..73H},
\cite{2014A&A...561L...8P}, \cite{2018A&A...620A.137R}.
In this method, for two filters e.g. $B$ and $V$,  the $B$ and
$V$ data points obtained in the same night through the same apertures are
plotted in a $B$ vs. $V$ flux diagram (Figure~\ref{fig:fvg_bv}).
The important feature is that the flux variations follow a linear relation
with a slope $\Gamma$ given by the host-free AGN continuum.
In the flux-flux diagram the host galaxy -- including the contribution of line emission from the 
narrow line region (NLR) -- lies on the AGN slope somewhere toward its fainter end.
With knowledge of the host colors, i.e. the host flux ratios $B/V$ and $r/z$,
the FVG analysis yields the intersection of the AGN slope (blue lines in Figure~\ref{fig:fvg_bv})
with the host slope (red dotted lines in Figure~\ref{fig:fvg_bv}) and thus the host fluxes
in the four filters $BVrz$ marked by green stars in Figure~\ref{fig:fvg_bv} and listed in
Table~\ref{tab:host_values}.
Then the host SED template in Figure~\ref{fig:host_sed} is shifted vertically to fit
the $BVrz$ host fluxes.
Finally this SED allows to extrapolate the $JHK$
fluxes of the 3C\,273 host for our apertures.
The values are listed in Table~\ref{tab:host_values}.
Compared with the total $JHK$ fluxes (Table~\ref{tab:observations})
the host contributes between about 30\% in $J$ and 15\% in $K$.

\begin{table}
  \centering
  \caption{3C\,273 average host, AD and dust fluxes in mJy,\\ $\diamond\,=$
     host extrapolation (Figure~\ref{fig:host_sed}),\\
    $*\,=$ power law AD extrapolation (Figure~\ref{fig:ad_fit}).
    \label{tab:host_values} }
  \begin{tabular}{c cccc}
  \tableline
    Filter & Host  & AD  & Dust  \\ \hline
    $B$ & 1.00\,$\pm$0.1 & 22.04\,$\pm$ 1.39&  --\\
    $V$ & 3.00\,$\pm$0.3 & 21.34\,$\pm$ 1.12& -- \\
    $r$ & 4.00\,$\pm$0.3 & 19.42\,$\pm$ 1.41& -- \\
    $z$ & 7.30\,$\pm$0.3 & 17.29\,$\pm$ 1.57& -- \\
    $J$ & 8.87\,$\pm$0.5$^{\diamond}$ &  15.57\,$\pm$0.85$^{*}$ &  5.65\,$\pm$0.92 \\
    $H$ & 10.12\,$\pm$0.5$^{\diamond}$ & 14.05\,$\pm$1.05$^{*}$ & 17.21\,$\pm$1.99 \\
    $K$ & 10.01\,$\pm$0.5$^{\diamond}$&  12.86\,$\pm$1.15$^{*}$ & 53.85\,$\pm$3.42 \\
    \tableline
  \end{tabular}
\end{table}

\subsubsection{Accretion disk}
\label{sec:ad}

\begin{figure}
  \centering
  \includegraphics[width=\columnwidth]{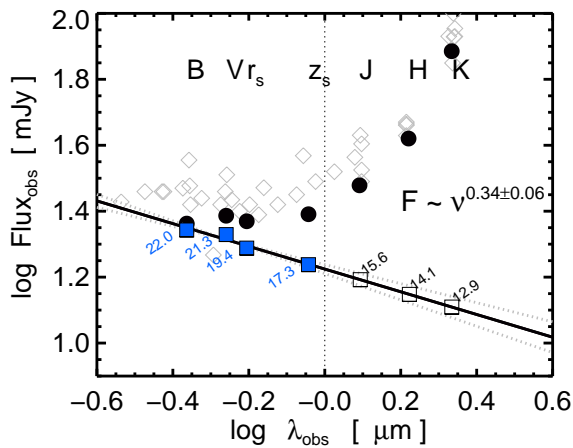}
  \caption{Derivation of the AD contribution to the NIR filters.
    Total average fluxes are plotted as black circles, host subtracted $BRrz$ fluxes as blue squares.
    The power law fit between $BRrz$ host subtracted
    fluxes $F\sim \nu^{\alpha}$ with \hbox{$\alpha = 0.34\pm0.06$}
    is plotted as a solid black line, fit error with dashed grey lines and $JHK$ interpolated values
    as open squares, the values for the AD fluxes are labeled. Additional photometry from the
    NED (https://ned.ipac.caltech.edu/)
     is plotted in the background
    as grey open diamonds.\label{fig:ad_fit}
  }
  \vspace{3mm}
\end{figure}

To estimate the spectrum of the AD in $BVrz$,
we subtracted the host contribution  (Table~\ref{tab:host_values})
from the mean total fluxes (Table~\ref{tab:observations}).
The result is shown as blue squares in Figure~\ref{fig:ad_fit}.
The power law fit to the $BVrz$ data points yields $F_{\nu} \sim \nu^{\alpha}$,  with $\alpha = 0.34\pm0.06$,
in agreement with the spectral index $\alpha\,=\,+1/3 $ found by \cite{2008Natur.454..492K} for six quasars.
Their study of the NIR component of the AD as seen in polarized light
reveals that the AD spectrum continues towards the NIR
with the same power-law slope as measured at optical wavelengths.
Adopting that this holds also for 3C\,273, we take the
AD contribution to the NIR bands from the power law fit with values as labeled at the open squares
in Figure~\ref{fig:ad_fit} and listed in Table~\ref{tab:host_values}.
The AD contribution to the total NIR fluxes is $J\sim50\%$, $H\sim30\%$, and $K\sim15\%$.

\subsubsection{Dust light curves}
\label{sec:dust_lc}

We derived the dust light curves from the observed $JHK$ light curves
(Fig.~\ref{fig:optical_lc_all}) by subtraction of both
the $JHK$ host galaxy contribution (Tab.~\ref{tab:host_values})
and a suitably scaled light curve of the AD.
For this AD light curve we used
the flux-scaled host-subtracted $V-$band light curve $LC(V$).
The scaling factor, $SF$, was determined from the power-law AD extrapolation,
e.g. in the $J-$band with values from Table~\ref{tab:host_values}:
$SF(J) = F_{\rm AD}(J)/F_{\rm AD}(V) = 15.57 / 21.34$.
This yields at the $JHK$ bands, respectively,

$$
  LC(dust) = LC(total) - F(host) - SF \times LC(V) \label{eq:lc}
$$

\begin{figure}
  \centering
  \includegraphics[width=0.95\columnwidth]{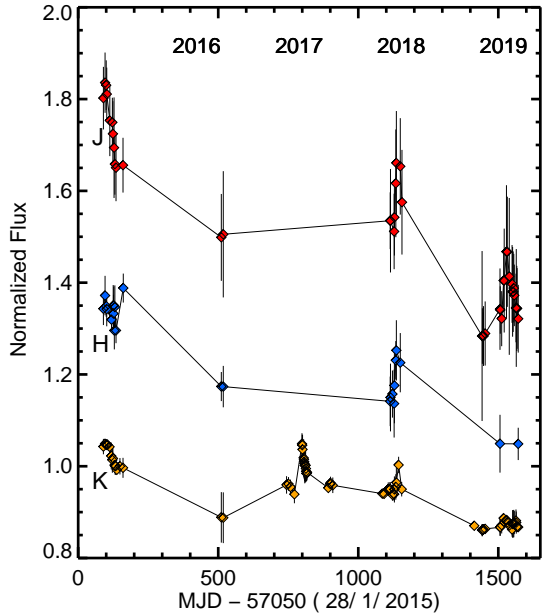}
  \caption{Normalized $JHK$ ''pure dust'' light curves, after subtraction of host galaxy and AD contribution.\label{fig:lc_dust}}
  \vspace{3mm}
\end{figure}

The resulting dust light curves are shown in Figure~\ref{fig:lc_dust}.
Compared to the observed NIR light curves in Fig.~\ref{fig:optical_lc_all} we find the following changes:
\begin{itemize}
\item
  In $K$ the shape of the light curve is similar, but the amplitude increases from about 12\% to about 20\%.
\item
  In $H$ the decrease between 2015 and 2016 becomes more pronounced and the amplitude increases
  from about 15\% to about 30\%.
\item
  In $J$ the shape of the light curve changed; the increase between 2015 and 2016 reverses now to a decrease,
  similar to what is seen in the $H$ and $K$ bands. The amplitude increases from about 10\% to about 40\%.
\end{itemize}
To summarize, compared to the observed NIR light curves,
the dust light curves show more coherent variations and stronger amplitudes.

\begin{figure*}
  \centering
  \includegraphics[width=0.6\columnwidth]{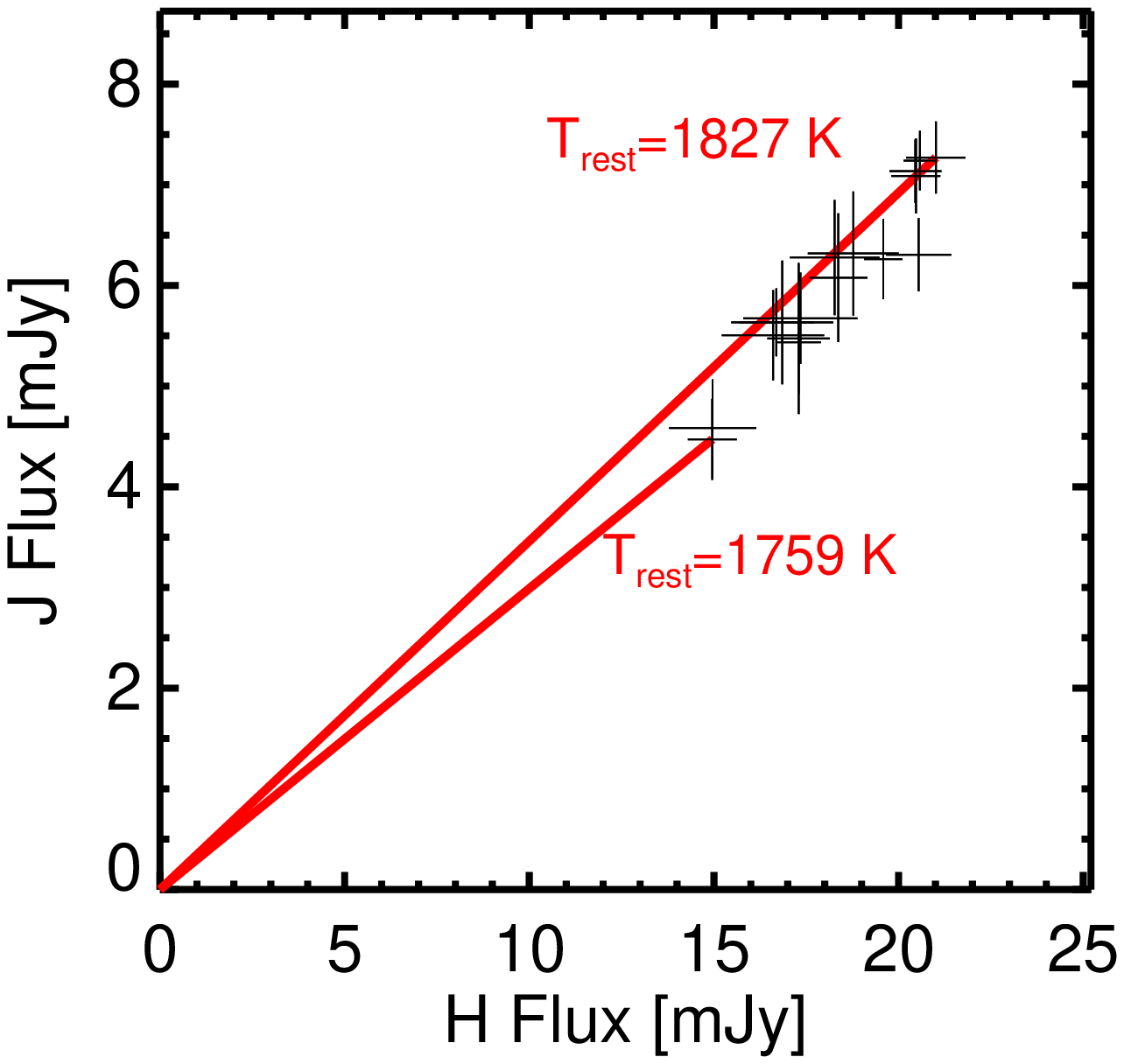}
  \includegraphics[width=0.6\columnwidth]{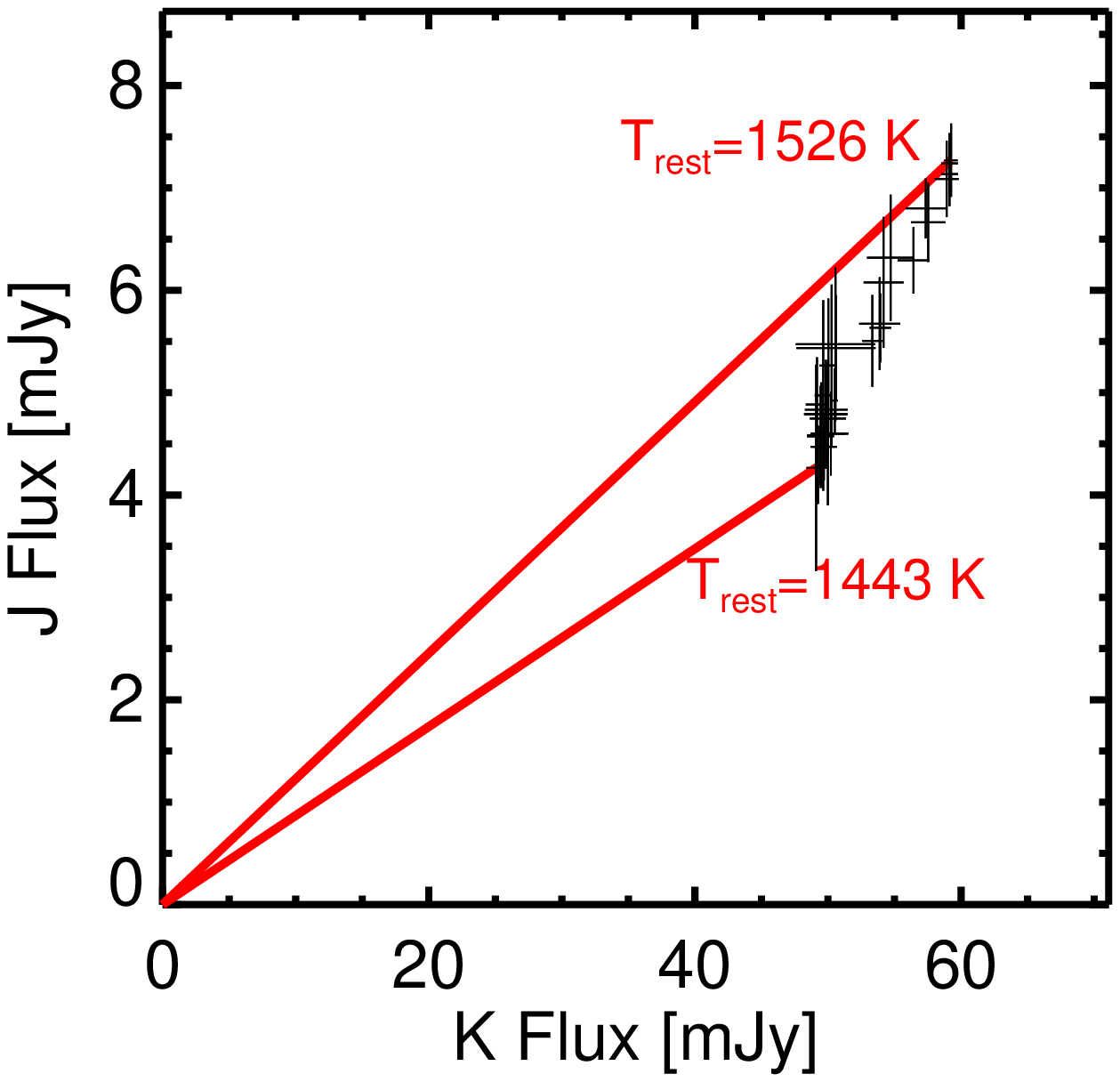}
  \includegraphics[width=0.6\columnwidth]{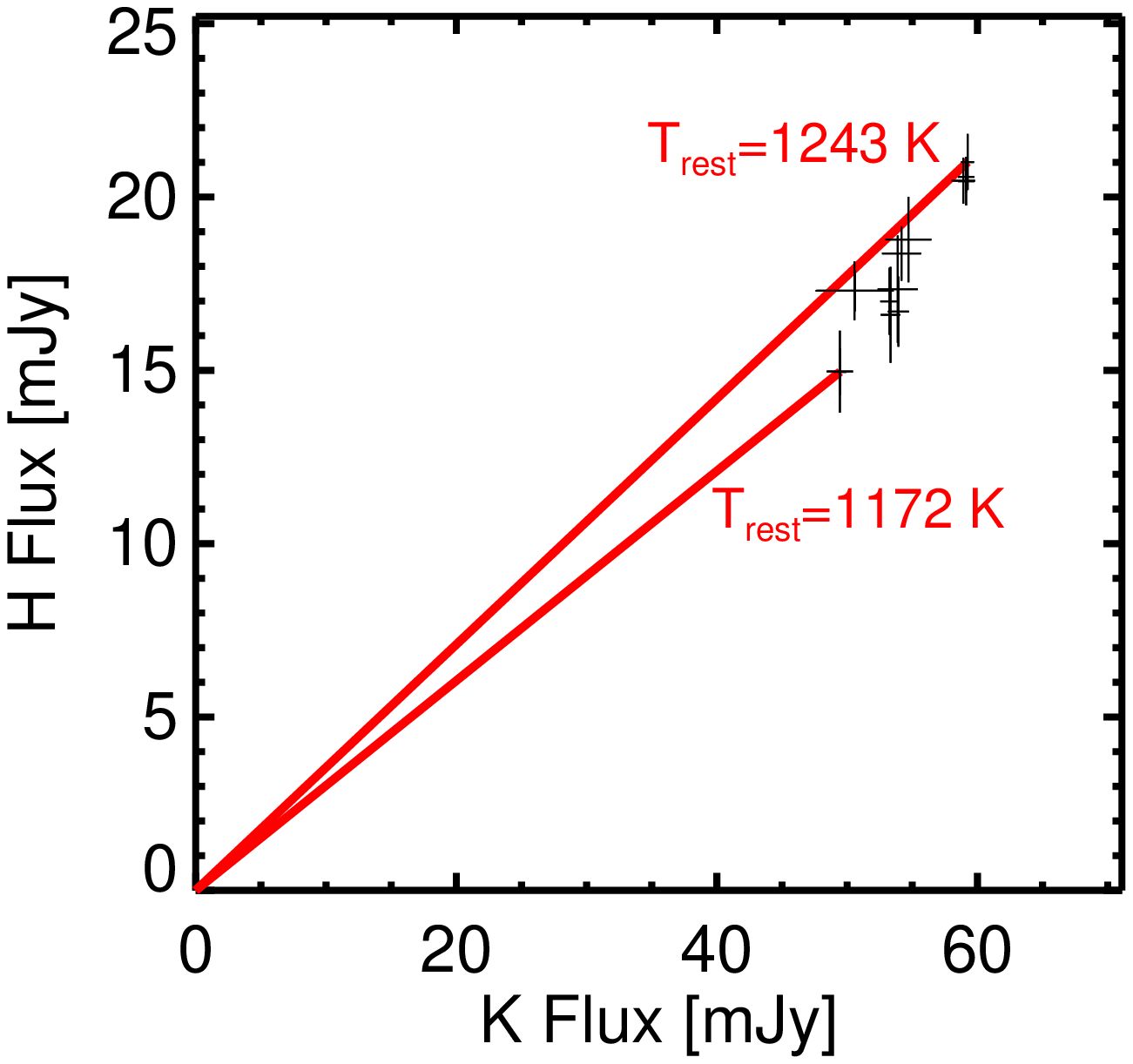}
  \caption{Flux-flux diagrams for $JH$ (left), $JK$ (middle) and $HK$ (right)-bands
    (where the AD and host contribution has been subtracted). Red lines show the faint
    and bright state, labeled with the corresponding blackbody temperature in the rest frame.
    \label{fig:fvg_nir_noadnohost}
  }
  \vspace{3mm}
\end{figure*}

\subsection{Nature of the dust variability}
\label{sec:dust_de}

Now we examine the variability
properties of the pure dust emission in the NIR, after subtraction of the host
and AD contribution. Figure~\ref{fig:fvg_nir_noadnohost} shows the
flux-flux diagrams for the $JHK$ filter pairs.
For all pairs ($J/H$, $J/K$, $H/K$)  the variations  are correlated.
This adds confidence that the creation of the dust light curves
from the observed NIR light curves by means of subtraction of the host
and AD contribution is sound.

The thick red lines in Figure~\ref{fig:fvg_nir_noadnohost} mark the range of color
temperatures between the bright and faint states,
calculated
for Planckian curves
in the rest frame of 3C\,273.
We make the reasonable assumption that the dust grains are at a mix of temperatures.
Then the shorter wavelength filters are more sensitive to the hotter dust 
grains. 
This explains the range of measured color temperatures between \hbox{1200\,K} and \hbox{1800\,K.}
This range is consistent with expected dust temperatures.
For comparison, the sublimation temperatures $T_{\rm sub}$ of graphite dust grains are estimated
to be $1500-1900$\,K \citep{Barvainis87, 2007A&A...476..713K}.

For all filter pairs, the color temperatures change by about 5\% (i.e. a factor 1.05)
between the bright and faint states.
For Planckian curves the luminosity is proportional to the 4th power of the
temperature ($L \propto T^4$). With this assumption\footnote{This assumption
holds for dust grains with diameter $a$ larger than the wavelength $\lambda$.
For smaller grains the dust emissivity properties come into play,
yielding up to $L \propto T^6$ {\bf for emissivity exponent $\beta=2$}, see e.g. \cite{2003pid..book.....K}.}
the amplitude of the dust luminosity is then $1.05^4 - 1 = 0.2$,
i.e. 20\%.
The amplitude of the $V-$band light curve,
i.e. amplitude of the triggering variations from the AD, is about 25\% (Fig. \,\ref{fig:lag_V_K}).
Thus the amplitude of the dust luminosity is a bit smaller than that of
the triggering variations from the AD. This is consistent with simple expectations that the
echo amplitude does not exceed the amplitude of the driving signal.

\begin{figure}
  \centering
  \includegraphics[width=0.9\columnwidth]{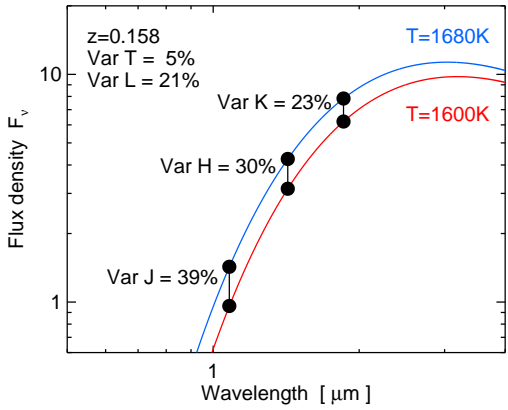}
  \caption{Illustration of the Wien tail amplification when measuring the variability of dust emission.
    The temperature $T$ varies by 5\%, raising from 1600 K to 1680 K.
    Then the dust luminosity $L$, integrated over the Planckian curves varies by 21\%.
    However, a measurement in the NIR filters at the Wien tail of the Planckian yields larger variabilities,
    which increase with decreasing wavelength from 23\% at $K$ to 39\% at $J$ (calculated in the rest frame
    for z=0.158).
    \label{fig:wien_amplification}
  }
  \vspace{3mm}
\end{figure}

The amplitudes are about $(60-50)/55 = 0.18$ in $K$,
$(19-14)/16.5 = 0.3$ in $H$, and
$(7.75-5.25)/6.5 = 0.39$ in $J$.
With decreasing wavelength, the amplitudes of the dust light curves
increase and exceed the amplitude of the driving signal.
As explanation for the different $JHK$ amplitudes we suggest
that the filters measure the dust emission on the
Wien tail of the Planck function.
The sensitivity to temperature changes increases toward shorter wavelengths.
This is illustrated in Figure~\ref{fig:wien_amplification}.
While a priori the echo amplitude is not expected to exceed the amplitude of the driving signal,
we here encounter the  case of an amplitude amplification
which we call Wien tail amplitude amplification.
Because of this amplitude amplification,
even in $K$ the amplitude of 0.18 may be an overestimate of the echo amplitude of the luminosity;
this may become relevant for the light curve modeling in Section~\ref{sec:bowl_shaped}.

In the following we will use the dust light curves as derived in $JHK$
from the observed NIR light curves by means of subtraction of the host
and AD contribution and adopt that the variations are essentially
caused by a change of the mean dust temperatures.


\subsection{Cross correlation analysis}
\label{sec:time_lag}

We determined the time lag of the dust variations (echo) against the AD variations (driving signal) by different methods and by direct inspection of the time-shifted lightcurves. For the AD we use the combination of the host subtracted $V$-band light curve from \cite{2017ApJS..229...21X}, \cite{2019ApJ...876...49Z} and the one in this work.

The cross correlation functions (CCF) yield the average flux-weighted time lag \citep{1991ApJS...75..719K, 1991vagn.conf..343P}. Our dust light curves are rather sparse. Therefore, we apply the discrete correlation function (DCF) by \cite{1988ApJ...333..646E}, which has been designed for sparse and unevenly sampled light curves.
We also applied the Z-transformed DCF (ZDCF) which is known to provide more conservative, larger error estimates \citep{Alexander1997}. 
We also applied the interpolated cross-correlation (ICCF), introduced by \cite{1986ApJ...305..175G}. Its application has proven to work well, if the light curves are well sampled, as is the case at least for the $V$-band light curve.
Additionally we use the von Neumann mean-square estimator for reverberation mapping data (VNRM) introduced by \cite{2017ApJ...844..146C}.

The DCF and ICCF centroids are calculated where the correlation value $r$ is $r > 0.8*r_{max}$. For the VNRM, the lag corresponds to the minimum of the VNRM estimator and for the ZDCF to the maximal likelihood. The estimation of the lag uncertainty in the DCF, ICCF and VNRM is calculated via the flux randomisation/random subset selection method (FR/RSS) by \cite{1998PASP..110..660P}, here applied to 2000 modified light curves. In case of the ZDCF error estimation, the default parameters were used.

\subsubsection{Time lag of the dust emission}

\begin{figure}
\centering
\includegraphics[width=0.99\columnwidth]{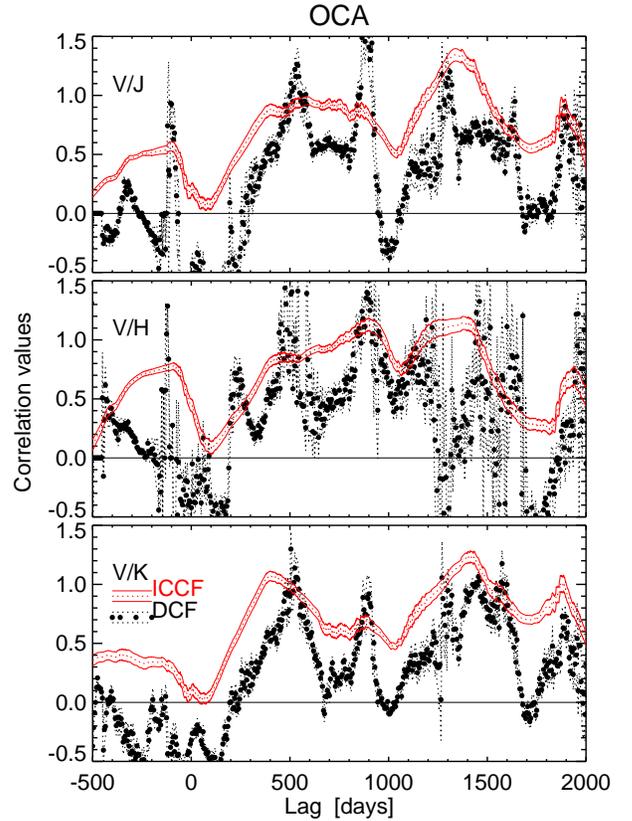}
\caption{DCF (circles+pointed lines) and ICCF (red pointed+solid lines) between V and $JHK$ dust light curves for our NIR observing campaing. The ICCF is shifted up by 0.5. The filter combinations are $V/J$ (top), $V/H$ (middle) and $V/K$ (bottom). \label{fig:dcf_V_J_K}}
\end{figure}

Figure~\ref{fig:dcf_V_J_K} shows the DCF together with the ICCF for the three filter combinations V/J (top) V/H (middle) and V/K (bottom), the ICCF is shifted up by 0.5. The DCFs in Figure~\ref{fig:dcf_V_J_K} show two prominent peaks at $\sim$\,500\,d and $\sim$\,850\,d for the three filters. Also the lag range between 1300-1500\,d shows a correlation value larger than 0.5 (noisy for $H$-band), the long lag peak at $\sim$\,1400\,d is also present in the ICCF for the three filters. For $J$ and $H$ , the ICCF does not show the two main peaks (at $\sim$\,500\,d and $\sim$\,850\,d), rather they are smooth together, which points out the problem of the interpolation when one of the light curves (trigger or echo) has long gaps.
Our $J$ and $H$ light curves are  poorly sampled and one observation season (2017, see Figure~\ref{fig:lc_dust}) with a pronounced turn-up of the $K$-band light curve is missing in $J$ and $H$. For the $K$-band, the peak at $\sim$\,850\,d essentially disappears (in the ICCF).
 Figure~\ref{fig:ccf_appendix_oca} in Appendix~\ref{sec:appendix} presents all the CCF obtain with the different methods for the three filters, showing that all methods are consistent with each other.

\begin{figure}
\centering
\includegraphics[width=0.99\columnwidth]{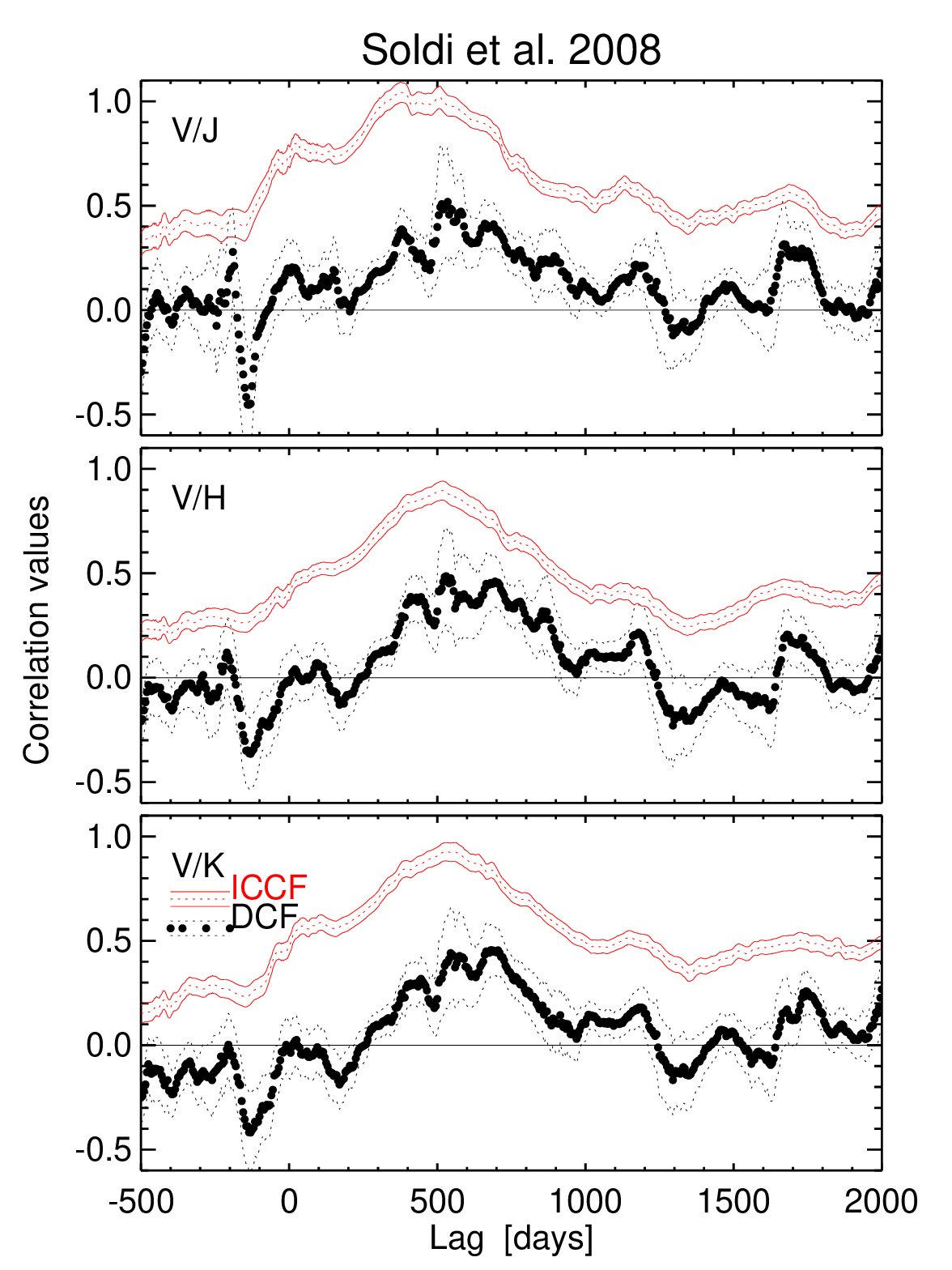}
\caption{Same as Figure~\ref{fig:dcf_V_J_K} but for \cite{2008A&A...486..411S} $V$ and NIR light curves (after AD subtracion).\label{fig:dcf_iccf_geneve}}
\end{figure}

Since our NIR campaign is only 1500\,d long, we are not able to confirm/reject lags on this long time scale. In order to check whether the long time lags are real ($\sim$ 800\,d and $\sim$ 1500\,d present in the CCF in Figure~\ref{fig:dcf_V_J_K}), we make use of the 30 years long light curves collected by \cite{2008A&A...486..411S}. 
We subtract the AD contribution in $JHK$, using the V-band light curve minus host (6\% of the average V band flux, \cite{2013ApJ...767..149B},Table~12) and $F_{\nu} \propto \nu^{0.34}$ (as for the OCA data, see Fig.~\ref{fig:ad_fit}), and compute the DCF, ICCF, ZDCF and VNRM between the dust light curves and the $V$-band light curve. The results for the DCF and ICCF are shown in Figure~\ref{fig:dcf_iccf_geneve}, where the ICCF is shifted up by 0.5. All filters show a maximal correlation value of $\sim$~0.5 located between 300 and 700\,d. None of these correlations shows any evidence favouring the long delays at about 800 and 1400\,d (see also Figure~\ref{fig:ccf_appendix_soldi} in Appendix~\ref{sec:appendix} for all the CCF in the three filters).
We found an average time delay for $JHK$ in the observer frame using the DCF $\tau = 600 \pm 60$\,d, ICCF $\tau = 550 \pm 50$\,d, ZDCF $\tau = 510 \pm 120$\,d and VNRM $\tau = 520 \pm 30$\,d. The lag values are slightly larger than the one reported in their study ($\tau_{\rm obs} \sim$\,420$\pm$\,84\,d.; $\tau_{\rm rest} \sim$\,365$\pm$\,73\,d.). However, their shorter lags can be explained by the contribution of the AD autocorrelation. This contribution is wavelength dependent and shifts the cross correlation values to smaller lags.
The CCF shapes and the time delays found for \cite{2008A&A...486..411S} data leads us to conclude, that for our OCA campaign, the lag has to be searched between 200 and 800 days.

The time lags found for our NIR observing campaign are listed in Table~\ref{tab:corr_values}.
For all three filters $JHK$, the time lags obtained via the DCF are consistent with each other within the errors.
For the ICCF, the $J$ and $H$ lag values show very large errors, hence
are less trustable. Even worse, the two main correlation
peaks (at 500 d and 850 d) are merged together (see Figure~\ref{fig:dcf_V_J_K}),
a fact which could be explained by
the interpolation of the 
$J$ and $H$ light curves across the 
gap between 2016 and 2018. For the $K$-band correlation the ICCF lag
(420 d) is smaller  compared to the DCF
($\sim$\,510\,d) and ZDCF ($\sim$\,550\,d.) but  agrees  
with the VNRM ($\sim$\,410\,d). For the best
sampled dust light curve, $K$-band, the $V/K$ correlation lies between 400 and 550 days taking all the CCF methods, with an average delay of $\sim$\,475\,days.

\begin{figure}
  \centering
  \includegraphics[width=\columnwidth]{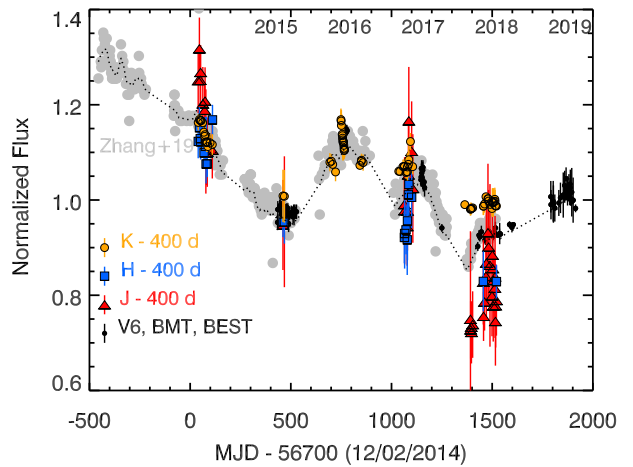}
  \caption{Normalized $JHK$ dust light curves back-shifted by 400 d
    and superimposed on the $V-$band light curve.
    The dust light curves basically match the $V-$band variation features,
    apart from the large dust amplitudes  in $H$ and $J$
    (caused by the Wien tail amplitude amplification, Sect.~\ref{sec:dust_de}).
    \label{fig:lag_V_K}
  }
\end{figure}

The light curves show clear variation patterns, allowing us to check visually via
back-shifting whether the different lag estimates appear consistent with the data.
Figure~\ref{fig:lag_V_K} shows the overlay of the AD and the back-shifted dust light curves,
with a back-shift of 400\,d.
In fact, the variation patterns match well but a spread or tolerance of $\sim 100\,$d
for the back-shift should be adopted.
A visually determined lag of 400 $\pm$ 100\,d is consistent with the
broad cross correlation function and the time lag calculations from Table~\ref{tab:corr_values}.

The rest wavelengths of $JHK$ are 1.08, 1.42 and 1.86\,$\mu$m.
The data do not indicate a significant
trend of a lag shortening with decreasing wavelength (only in case of the ZDCF, but with large errors specially in $J$),
as has sometimes been reported, e.g. \cite{2006ApJ...652L..13T}.
Thus, our data of 3C\,273 are in line with the relative wavelength independence of NIR lags
reported by \cite{2015OAP....28..175O}, which they attribute to a specific geometry for the dust.

In our data the $K$-band lag is more reliable than the $J$ and $H$ band lags due to a better time sampling, and likewise in \cite{2008A&A...486..411S} the uncertainty of the AD subtraction is larger in $J$ and $H$ than in $K$. Therefore, for both data sets we adopt the $K$-band lag as the optimal time lag for the hot dust.
Table~\ref{tab:corr_values} lists the $K$-band lag in days obtained with the different methods for the Soldi et al. light curves (fourth column). The lags for this work are in general shorter, only the ZDCF shows a slightly longer lag in the OCA campaign, but its error is high.
We take as final delay for the hot dust the average time lag in $K$ from the four CCF methods, because the results are consistent with each other within the errors. 
Table~\ref{tab:corr_values} shows that for this work we obtain a $K$-delay of $\tau_{\rm K, obs} = 474^{+24}_{-70}$\,d, which in the rest-frame corresponds to $\tau_{\rm K, rest} = 409^{+21}_{-61}$\,d. The average lag of the OCA observation is about 10\% shorter than that found for the \cite{2008A&A...486..411S} data. Our shorter delays may be explained by the fact that the 3C\,273 luminosity during the OCA campaign is about 28\% lower than during the 30 years before.\footnote{The host subtracted $V$-band flux for this work is AD$_{\rm OCA}$ = F$_{\rm total}$ - F$_{\rm host}\,\sim$ 21.3\,mJy (Table~\ref{tab:host_values}). For Soldi et al. it is about 27.5\,mJy, obtained from F$_{\rm total} = $29-30\,mJy (their Table~1) and subtracting 6\% host contribution (\cite{2013ApJ...767..149B},Table~12). Then the flux ratio is AD$_{\rm Soldi+2008}$ / AD$_{\rm OCA}\,\sim$\,1.28.}

\begin{table}
\centering
\caption{Observer's frame time delay in days between $V$ and $JHK$ dust light curves 
    from our campaign and from Soldi et al. (2008), after subtraction of the AD contribution.}
\label{tab:corr_values}
\begin{tabular}{c c c c c}
\tableline
Method & $V/J$& $V/H$ & $V/K$ & V/K (Soldi+08) \\  \tableline
DCF$_{\rm cen}$   & $532^{+22}_{-25}$   & $506^{+33}_{-45}$   & $513^{+12}_{-13}$ & $594^{+61}_{-69}$  \\ 
ICCF$_{\rm cen}$  & $549^{+209}_{-135}$ & $769^{+8}_{-219}$   & $420^{+36}_{-20}$ & $506^{+48}_{-49}$ \\ 
ZDCF              & $327^{+209}_{-14}$  & $462^{+75}_{-14}$   & $554^{+7}_{-170}$ & $516 \pm 128$  \\ 
VNRM              & $503^{+120}_{-195}$ & $564^{+120}_{-120}$ & $409^{+41}_{-80}$ & $519^{+54}_{-41}$ \\ [1pt] \tableline 
Average           &  $477^{+140}_{-90}$&  $575^{+59}_{-100}$  & $474^{+24}_{-70}$ & $534^{+62}_{-61}$  \\[1pt]   \tableline 
\end{tabular}
\end{table}

\subsubsection{Possible asymmetry of the cross correlation}

In the limit of infinite sampling, the CCF between the driving signal and its echo is equivalent to the convolution of the transfer function (TF) with the autocorrelation function (ACF) of the driving signal. Thus, the CCF may reveal higher order moments, e.g. asymmetries, of the transfer function. 

\begin{figure*}
\centering
\includegraphics[width=2.\columnwidth]{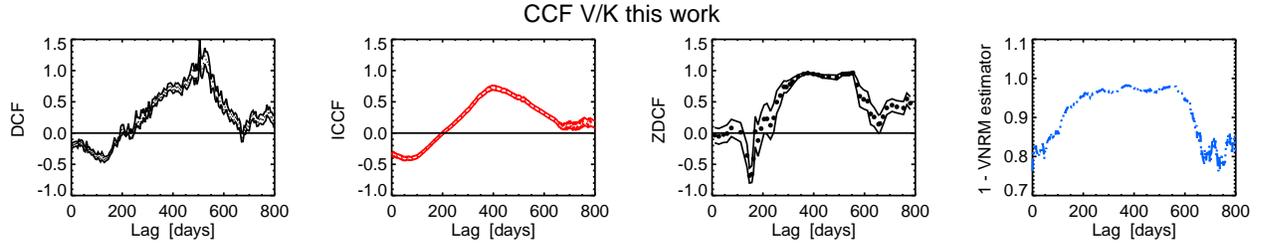}\\
\caption{From left to right: DCF, ICCF, ZDCF and VNRM estimator between $V$ and $K$ light curves. The zero correlation for DCF, ICCF and ZDCF is marked as an horizontal line.}
\label{fig:all_ccf}
\end{figure*}

\begin{figure*}
\centering
\includegraphics[width=2.\columnwidth]{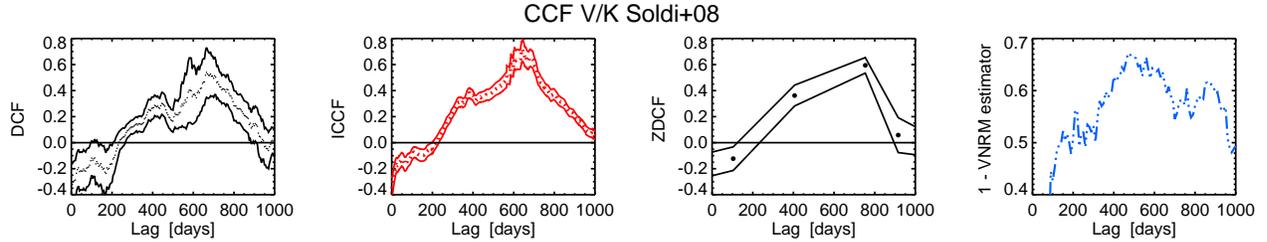}
\caption{Same as Figure~\ref{fig:all_ccf} for the $V$ and $K$ light curves in \cite{2008A&A...486..411S}, observations taken between years 1984-1994 due to a better sampling. }
\label{fig:all_ccf_soldi}
\end{figure*}

Figure~\ref{fig:all_ccf} shows the DCF, ICFF, ZDCF and VNRM estimator between $V$ and $K$ light curves within the time range of interest, until 800 days. All CCF show a broad correlation between 200 and 600 days. The DCF exhibit an interesting asymmetry: a peak at around 500-550\,d with a steep decline towards longer delays and a broad shoulder to shorter delays down to about 250\,d. On the other hand, the ICCF does not show a peak at 500--550\,d, but at $\sim$ 400 days and the correlation is more symmetric. The ZDCF and VNRM estimator show broader correlations but also a steep decline between $\sim$ 550\,d and $\sim$ 600\,d.

We also inspected the CCF shape using the best sampled part of the 
\cite{2008A&A...486..411S} light curves between
January 1984 and December 1994 with a median $V$ and NIR sampling of
around 7 and 22 days, respectively.
The computed CCF are shown in Figure~\ref{fig:all_ccf_soldi}.
For all the CCF methods, the cross correlations show a similar asymmetric shape.  The DCF and ICCF show a peak at about 700\,d, a steep decline to longer lags reaching the zero level at about 900\,d and a shallow tail reaching the zero level at about 200\,d; this shallow tail even reveals mildly a secondary peak at about 350\,d. In case of the VNRM estimator the asymmetry and the secondary peak are also present, but their are located at shorter lags, at around 200\,d and 500\,d respectevely. We come back to this asymmetric CCF shape in Section~\ref{sec:bowl_shaped}.

Finally we mention that there exists some differences in the CCF shapes
when using all observation data of \cite{2008A&A...486..411S}
(30 years) and when using a
part of these observations (11 years).
These CCF differences may point to the presence of anomalous responses
of the echo to the continuum variations  
as reported by \cite{2019arXiv190906275G} for the
H$\beta$ BLR of the Seyfert-1 NGC 5548 and a sample of other AGN including the PG quasars. As discussed by \cite{2019arXiv190906275G} such anomalies  best show up when inspecting the light curves, and they could be caused by, e.g., anisotropic continuum emission or absorbing clouds. Nevertheless, here for the dust reverberation of 3C273, we are not going further into these details.

\subsection{Dust Covering Factor}
\label{sec:dust_covering_factor}

The UV to NIR SED allows us to estimate the covering factor ($CF$) of the hot dust.
If the dust grains completely re-emit the absorbed UV radiation in the infrared,
then the covering factor is defined as
$CF = \Omega/4\pi = L_{\rm IR}/L_{\rm UV}$,
where $L_{\rm IR}$ is the total IR luminosity of the dust
and $L_{\rm UV}$ the total UV luminosity from the AD.
$CF$ can be approximated by the peak
luminosities \citep{2011MNRAS.414..218L}:

$$
  CF = 0.4 \times \frac{\nu_{\rm K} L_{\rm K}}{ \nu_{\rm UV} L_{\rm UV}}\label{eq:cf}
$$

For 3C\,273 we obtain $CF \sim 0.08$.
This agrees with the findings of \cite{2011MNRAS.414..218L} for a sample of 23 type-1 AGN,
where  $CF \sim 0.01 - 0.6$ with an average of $<CF> = 0.07 \pm 0.02$.

A small $CF$  suggests that the NIR dust emission originates in a  small angular range seen by the AD.
For a thin ring at $40^{\circ} < \theta < 45^{\circ}$  with $\theta$
measured against the equatorial plane, the dust covering fraction is about $CF =  0.06$.
We elaborate on this issue in Section~\ref{sec:bowl_shaped}.



\begin{figure*}
\centering
  \includegraphics[width=1.\columnwidth]{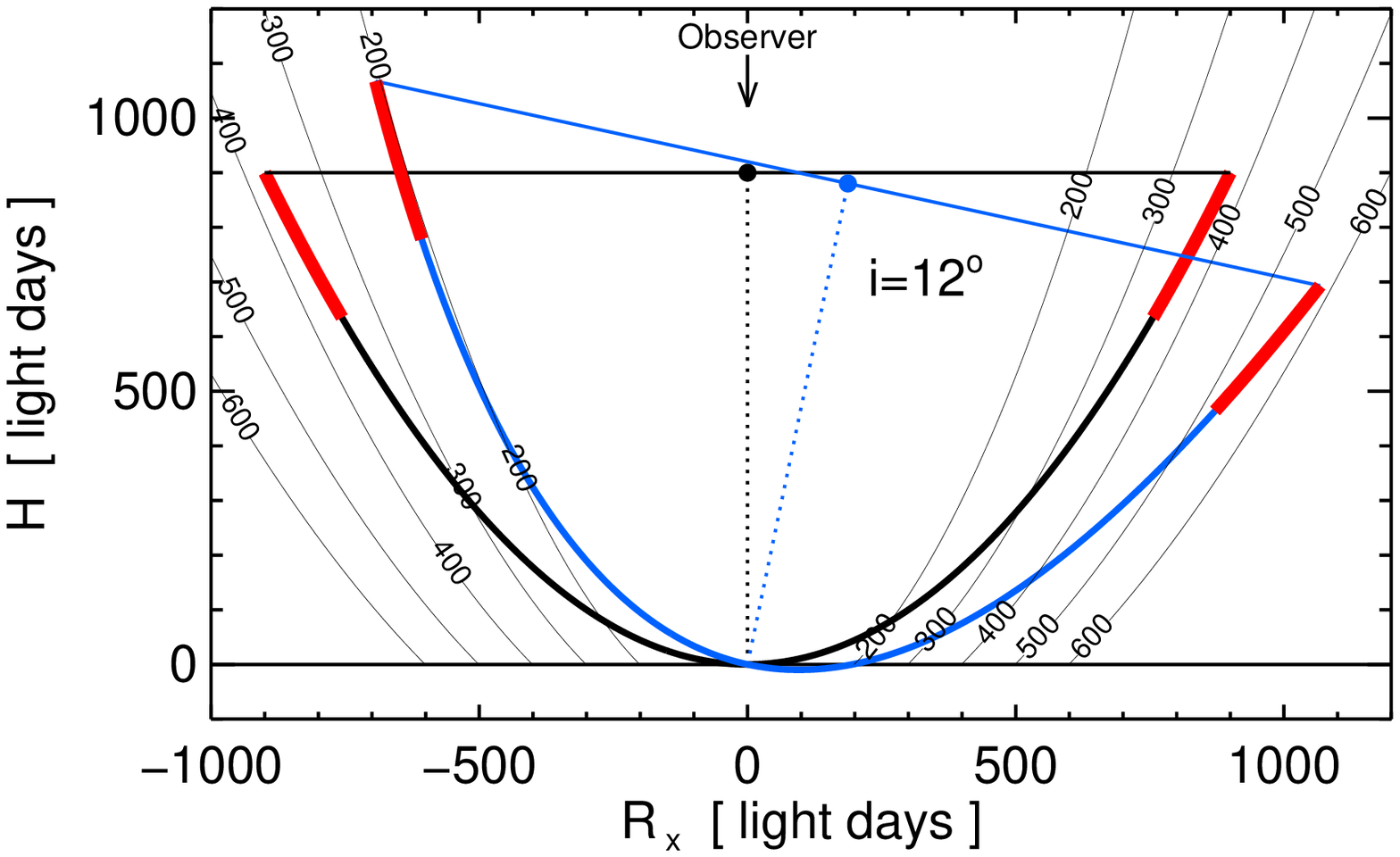}
  \includegraphics[width=0.85\columnwidth]{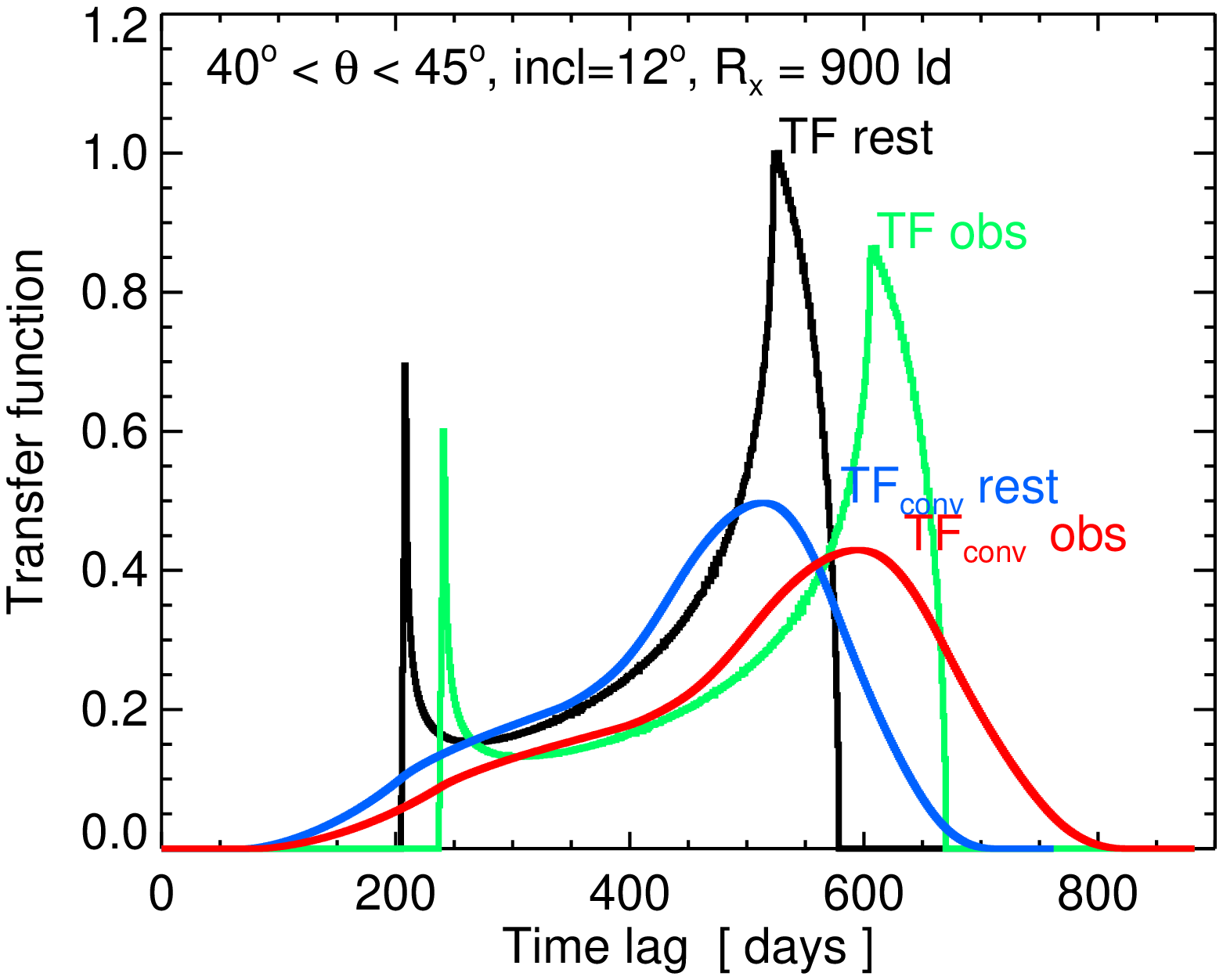}
  \caption{Left: Cuts through a bowl geometry with $R_x = 900$\,d.
    The face-on cut is plotted with a thick
    black line and a cut with an inclination $i=12^{\circ}$ with a thick blue line.
    The dust emission seen in the NIR is assumed to originate exclusively at the bowl edge
    marked with thick red line segments. The iso-delay contours are marked with thin black
    lines and labeled with the corresponding time lags. 
    In Section~\ref{sec:dust_geometry} we discuss this geometry 
    compared with a similar one shown in Figure~2 of \cite{2015OAP....28..175O}.
    Right: Transfer function (TF) for a bowl model with $R_x = 900$\,ld, $i=12^{\circ}$ and
    $40^\circ < \theta < 45^\circ$. Black = TF in the rest frame, green = TF shifted to the observers frame,
    blue = rest TF convolved with a triangle kernel of $300$\,d base-line, 
    red = convolved TF shifted to the observers frame.
    In Section~\ref{sec:tf1} we compare this TF with those shown in Figure~4 of \cite{2011ApJ...737..105K}.\\
    \label{fig:all_tfs}\label{fig:bowl_K_lc} }
\end{figure*}

\section{Bowl-shaped Geometry}
\label{sec:bowl_shaped}

\cite{2011A&A...527A.121K} performed $K-$band interferometric measurements of 3C\,273.
Modeling the visibility with a thin dust ring, they found an angular size of
$0.81 \pm 0.34$\,pc ($933 \pm 392$\,ld for the cosmology adopted here).
Recently, the \cite{2019arXiv191000593G} found a similar angular size of $0.28 \pm 0.03$\,mas
which -- adopting a Gaussian FWHM -- corresponds to a dust radius size of
$0.567 \pm 0.106$\,pc $= 675 \pm 126$\,ld and translates to ring radius of about 900\,ld. 
On the other hand, with RM technique we here obtain an average rest frame
time lag of $\sim\,410\,$d.
This is a factor two lower than expected, if both methods see
the same dust emission and if
the dust were located in the equatorial plane of the AGN.

Interferometry measures the projected size of the NIR emitting dust as seen from
the observer and does not take into
account the vertical structure of the dust. If the NIR emitting dust is not located
in the equatorial plane
but closer to the observer than the AD, then reverberation mapping
will produce a fore-shortening effect, i.e. yield about 2-3 times shorter time lags, as has been
discussed in \cite{2014A&A...561L...8P} (see their Figure.6)
and in \cite{2015OAP....28..175O} (see their Figures.2 and 3). 
In this section, we explore how far the difference between the interferometric
values and our RM value can be explained by a special geometry of the
dust emitting zone.

We here consider a paraboloidal bowl geometry, following \cite{2012MNRAS.426.3086G}.
The height $H$ of the bowl rim is $H \propto R_{\rm x}^{\rm 2}$,
with $R_{\rm x}$ being the radius  in the equatorial plane.
First, we adopt a half-opening angle \hbox{$\theta = 45^{\circ}$}
statistically justified by the fraction of type-1 to type-2
AGN \citep{1992ApJ...393...90H, 1989ApJ...336..606B},
and an inclination $i=12^{\circ}$ based on the orientation of the radio jet axis
against the line of sight to the observer \citep{2001Sci...294..128L, 2006A&A...446...71S, 2017ApJ...846...98J}.

Figure~\ref{fig:bowl_K_lc} (left)
shows such a bowl geometry with an equatorial radius of $R_x = 900$\,ld, 
as the reported interferometric dust ring radius.
A face-on bowl is plotted as a thick black line and a bowl with $i=12^{\circ}$  is plotted as thick blue lines.
Since the dust covering fraction $CF$ found in Section~\ref{sec:dust_covering_factor} is very low, we assume a small area where the dust emission occurs. As proposed by \cite{2012MNRAS.426.3086G}, it is located on the edge of the bowl-shaped dust torus, between $40^\circ < \theta < 45^\circ$ and marked with thick red lines (with $\theta$ being measured against the equatorial plane). The iso-delay contours are marked with thin black lines and labeled with the corresponding time lags. 

Note that the complete model is actually a bi-conical bowl model but
that in this model, the observer only sees the front side of the bowl; the back side below the equatorial plane is hidden (i.e. highly absorbed).

\subsection{Transfer function for the bowl-shaped geometry}
\label{sec:tf1}

We calculated the transfer functions (TF) for different bowl
sizes with an inclination angle of $12^{\circ}$ and an emission
range of the dust rim of $40^\circ < \theta < 45^\circ$.

The TF was calculated as follows: we sampled the dust rim in 3D space as
seen from the observer to a grid of 1\,ld
cell size, computed  for each cell the lag and calculated the TF as histogram over the cells.
While this TF is just a geometric approximation and does not account for clumpy dust structures and possible shadowing of dust clouds, it allows us to draw basic conclusions.

Figure~\ref{fig:all_tfs} (right) shows an example of the TF for a fixed bowl size of 900\,ld and a dust emitting region $40^\circ < \theta < 45^\circ$. It shows the TF in the rest frame (black) and redshifted  ($z=0.158$) to the observer's frame (green).

The TF shows a pronounced double-horned profile.
For comparison of our TF at inclination $12^{\circ}$
with the more sophisticated TFs calculated by \cite{2011ApJ...737..105K}, we refer to their
Figure~4, which shows the similarly double-horned TF of an optically-thin
torus at inclination $25^{\circ}$.
\cite{2011ApJ...737..105K} performed a clumpy torus calculation and presented
a plenty of TF details for different torus sections,
e.g. waning effect, shielding of clumps (optically thickness),
and even how much the observer may see from the torus back side,
i.e. from below the equatorial plane.
In our observational paper here we skip these details and 
continue the analysis with the geometric TF shown in Fig.~\ref{fig:all_tfs}.

As mentioned above (Section~\ref{sec:time_lag}), the cross correlation
between the driving signal and its echo is equivalent
to the convolution of the TF with the autocorrelation function (ACF) of the driving signal.
In Figure~\ref{fig:all_tfs}, the convolution of the TF (rest) with a kernel is also shown.
The results are  similar for different kernel shapes and widths,
hence quite stable against the details of the kernel.
We here used a triangular kernel of $300\,$d baseline (in the rest frame) as an approximation,
which was derived from  the autocorrelation of the $V-$band light curve shown in \cite{2019ApJ...876...49Z},
their Figure\,4. The resulting convolved TF$_{\rm conv}$ in the rest frame is plotted in blue, and in red in the observers frame.
The convolved TF$_{\rm conv}$ (obs) has a broad peak at about 550~d and a short-$\tau$ tail reaching to about 200~d (Fig.~\ref{fig:all_tfs}, right). This asymmetric TF matches some of the observed CCFs remarkably well, e.g. the DCF from the OCA campaign shown in 
Figure~\ref{fig:all_ccf} and the DCF, ICCF and ZDCF from 
Soldi et al.'s long campaign shown in Fig.~\ref{fig:all_ccf_soldi}.

\subsection{Modelling the dust echo light curves}
\label{sec:model_lc}
\begin{figure}
  \centering
  \includegraphics[width=0.9\columnwidth]{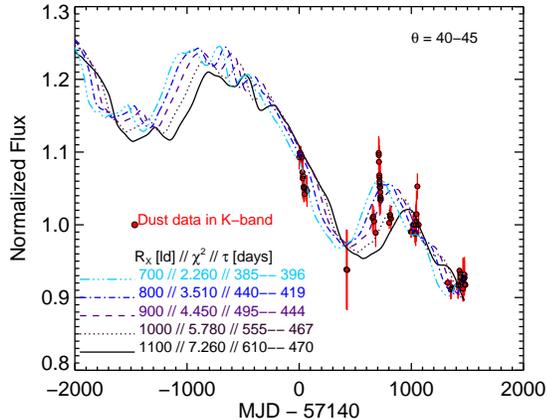}
  \caption{Echo light curves for different bowl sizes $R_x$.
    $i=12^\circ $ and $40^\circ < \theta < 45^\circ$.
    The lines represent the signal light curve convolved with different $TF_{\rm obs}$.   
    Red points show the dust light curve in the $K-$band. 
    The parameters of the bowl models
    are labeled.\\
    \label{fig:chi_bowl}
  }
\end{figure}
We checked the bowl-model
further using the light curves directly.
For the driving signal we used
the host-subtracted $V-$band light curve and to reduce high frequency noise,
we smoothed the signal light curve with a box car function (box size 100\,d).
We convolved the signal light curve with $TF_{\rm obs}$, 
the transfer function in the observers frame, yielding the modeled echo light curve.
Note that all calculations are made in the observers frame.
 
Figure~\ref{fig:chi_bowl} shows modeled echo light curves 
for some bowl sizes $R_x$ around the dust interferometric radius, between 700 and 1100\,ld.
Each model is plotted as a colored solid line and labeled in the inset table 
with the bowl size $R_x$, $\chi^2$, and the time lag 
(found via VNRM and DCF centroid) between the AD signal and the modeled echo light curves.
The echo models yield a large amplitude comparable to that of the signal light curve.

The best (i.e. smallest) $\chi^2$ is reached for \hbox{$R_x = 700$\,d} (Tab.~\ref{tab:bowl_param}). However, for this bowl size the average time lag \hbox{$\tau_{\rm R_x = 700} \sim 400$\,d} is shorter than the average observed \hbox{$\tau_{\rm K, obs} \sim 475$\,d} (Tab.~\ref{tab:corr_values}). On the other hand, a bowl size of \hbox{$R_x = 900$\,ld} yields the best lag agreement between model and data, while the $\chi^2$ values are not optimal. The $\chi^2$ values depend not only on the match of the lags but also on the match of the amplitudes. Because of the Wien tail amplification of the NIR dust light curves (Fig.~\ref{fig:wien_amplification}), we suggest that the $K-$band amplitude is too large to properly match even the best model.

\begin{table}
  \centering
  \caption{Summary of different bowl parameters with the corresponding average time
    lag $\tau_{\rm avg}$ (from DCF centroid and VNRM)
    and $\chi^2$ value of the fit to the $K-$band data.
    \label{tab:bowl_param}}
  \begin{tabular}{cccc}
    \tableline
    $R_x$ [ld]  &  $\theta$ [$^\circ$]  & $\tau_{\rm avg}$[days]  & $\chi^2$ \\ \tableline
    $700$  &  $\,40\,-\,45\,$  &  $390 \pm 10$  & $2.26$ \\
    $800$  &  $\,40\,-\,45\,$  &  $430 \pm 15$  & $3.51$ \\
    $900$  &  $\,40\,-\,45\,$  &  $470 \pm 40$  & $4.45$ \\ 
    $1000$ &  $\,40\,-\,45\,$  &  $510 \pm 60 $  &  $5.78$ \\ 
    $1100$ &  $\,40\,-\,45\,$  &  $540 \pm 100$  &  $7.26$\\ \tableline 
  \end{tabular}
\end{table}

Even with a relatively poor $\chi^2$, the average delays found for the echo light curves for bowl-sizes $R_{\rm x}\,=\,900\pm200$\,ld agree with the observed delay range found via CCF techniques (Table~\ref{tab:corr_values}).
The exact determination of the bowl model parameters requires more data. Nevertheless, the modeling leads us to conclude:
If the dust emission comes from an inclined ring above the equatorial plane of the AGN, it produces both a foreshortening effect and a large amplitude variation of the dust echo consistent with the observations.

\section{Discussion}\label{sec:discussion}

We found a hot dust lag in the $K$-band of \hbox{$\tau_{\rm  rest} \sim 410$\,d} for the luminous quasar 3C\,273, consistent with the lag \hbox{$\tau_{\rm rest}~\sim$~460 days} obtained using Soldi et al.'s 30 years long light curves (after AD subtraction) and taking into account that the luminosty during our campaign was about 25\% lower.
This dust lag is smaller than the predicted one of about 1000\,d from the lag -- luminosity relation with slope $\alpha = 0.5$ and the interferometry value of $\sim$~900\,light days. We here discuss the results and some implications.

\subsection{On the dust geometry}
\label{sec:dust_geometry}

In order to bring the observational constraints into a consistent picture,
we considered a bowl-shaped torus geometry as proposed by \cite{2012MNRAS.426.3086G},
where the dust emission originates from the edge of the bowl rim with a small covering angle
$40^\circ<\theta<45^\circ$, as justified by the small
$CF$  ($\theta$ is measured against the equatorial plane).
We used an inclination angle of $12^{\circ}$ indicated from radio jet
studies \citep{2001Sci...294..128L, 2006A&A...446...71S, 2017ApJ...846...98J}.
It is clear that the exact parameters of the bowl are not uniquely determined
and -- in the frame of this observational paper -- we
have to be restricted to some reasonable cases.
We also did not consider clumpy dust distributions; this should not affect our results
as long as the BLR shields the bulk of the dust (at $0^\circ < \theta < 40^\circ$) from heating by the AD.

For an inclined bowl model the (simple geometric) transfer function (TF)
is double-horned and yields an asymmetric cross correlation similar to that found in the data.
The convolution of the TF with the host-subtracted $V-$band light curve (as proxy for
triggering signal light curve) yields the echo light curve.
The modeled echo light curves for different equatorial sizes $R_x$ around the interferometry radius ($R_x = 900\pm200$\,ld) are in agreement with the observed $K-$band dust light curve, with the average time delay and with the CCF shape.

Oknyansky, Gaskell \& Shimanovskaya (2015) presented in their Figures~2 and~3 a \textit{dust-cone} geometry, where the walls of the cone coincide essentially with an isodelay surface. This ``OGS-model'' looks quite similar to that in Figure~\ref{fig:bowl_K_lc} here which is based on the model of \cite{2012MNRAS.426.3086G}. The difference between the two models is that OGS's \textit{dust-cone} reaches from the equatorial plane up to about $\theta=45^\circ$, thus has a much larger covering angle ($>30^\circ$) seen from the AD than the dust-gloriole in the ``GKR-model'' of Goad, Korista \& and Ruff (2012). 
To bring the covering angle of the \textit{dust-cone} into agreement with the small (8\%) dust covering fraction (Sect.~\ref{sec:dust_covering_factor}), \cite{2007arXiv0711.1025G} had already proposed a shielding of the dust-wall by, for instance, randomly distributed BLR clouds located between AD and dust-wall (their Fig.~10). Then, despite a large covering angle, the intensity of the AD's radiation field reaching the dust-wall is reduced by the absorption in the BLR, and this may lead in the net effect to the calculation of a small covering factor. We have checked whether the available data are able to distinguish between the two models. We calculated the geometric TF for the OGS-model in the same manner as for the GKR-model. For $i=12^\circ$ and R$_{x} = 900$~ld the TF is also double-horned and shows -- after convolution with a triangle kernel of 300~d baseline -- an asymmetric profile, similar to that of the GKR-model depicted in Figure~\ref{fig:all_tfs} (right). Because the two TFs are so similar, the current data will not allow us to distinguish between the two models. Likewise the current interferometry data are too sparse and uncertain to allow for discriminating between the dust-gloriole (sharp ring) and the dust-cone (smeared ring seen in projection). Note that both models yield the foreshortened lags.

As is sometimes the case, the reality may lie in a synthesis or mixture of the two models. Such a refined model could be a dust-cone whereby the density of BLR clouds located between AD and dust-wall decreases with increasing $\theta$. 
This results in a shielding of the wall which is large close to the equatorial plane and decreases towards the cone edge
at $\theta = 45^\circ$. 
This refined model can also be described as a 
gloriole-like cone with a short wall extension towards the equatorial plane (so that the covering angle becomes larger than the about 5 degree wide ring) and some BLR clouds located between AD and dust-wall producing sufficient extinction (so that the net resulting dust covering factor remains small). 
In the net effect, both model descriptions are equivalent.
While the final answer has to be left to the future, for simplicity we here will continue with the ``dust-gloriole'' model. 

Then the important conclusion is that the hot dust emission of 3C\,273
comes essentially from a gloriole-like inclined ring (or the upper part of a cone) 
located above the equatorial plane of the AGN.

Commonly the sublimation radius $R_{\rm sub}$ is estimated following
\cite{Barvainis87,2014ApJ...788..159K}
and, as pointed out by \cite{2007arXiv0711.1025G},
assuming that all UV photons from the AD reach the dust zone without absorption by BLR gas.
If the hot dust is essentially located at the bowl edge ($\theta \approx 45^\circ$),
then $R_{\rm sub}$,  i.e. the 3-dimensional distance of the dust from the AD,
becomes larger than $R_{\rm x}$, namely:
$R_{\rm sub} = R_{\rm x} / cos(\theta) \approx 1.4 \cdot R_{\rm x}$ for $\theta \sim 45^\circ$.
This may explain -- at least partly -- why  $R_{\rm sub}$ is larger than
the interferometric ring size $R_{\rm ring}$ measured for some sources by
\cite{2007A&A...476..713K} and \cite{2011A&A...527A.121K}.
Likewise,  with $R_{\tau} = c \cdot \tau$, one obtains
$R_{\tau} / R_{\rm x} = 1 / cos(\theta) - 1 \approx 0.4 $ because of
the geometric foreshortening effect of the reverberation signal.
Then $R_{\rm sub}$ is expected to be about a factor 1.4/0.4 = 3.5
larger than $R_{\tau}$.

Next we briefly address some alternative models.
\cite{2011A&A...525L...8C} and \cite{2017ApJ...846..154C} considered the origin of the BLR
and proposed the Dusty Outflow Model where the dust clouds are radiatively accelerated.
Likewise, \cite{2015OAP....28..175O} proposed that the hot dust emission
comes from the near side of a hollow bi-conical outflow.
Moreover, to explain the changing look AGN like NGC~2617 \cite{2018rnls.confE..12O}
proposed that occasionally swirling hot dust clouds populate even the AGN polar region.
For the Seyfert\,2 NGC1068, \cite{1993ApJ...409L...5B} and \cite{1993ApJ...419..136C} resolved  the mid-IR emission to be aligned with the [OIII] ionization cone, i.e. perpendicular to the dust torus plane. This is unexpected within AGN unified model \citep{1993ARA&A..31..473A}. \cite{2000AJ....120.2904B} explains this polar dust emission as a strongly beamed re-emission from the nuclear radiation. Based on interferometry, \cite{2013ApJ...771...87H} observed a polar mid-IR emission also for the Seyfert\,1 NGC\,3738 and they proposed that the polar dust may originate from a dusty wind which is driven by radiation within the hot region of the dust torus.


According to the AGN unified scheme the BLR should lie inside the dust torus.
We here check whether the published BLR lag measurements
of 3C\,273 are consistent with the rest frame dust lag of \hbox{$\tau_{\rm rest} \sim 410$\,d}  and a torus radius $R_{\rm x} \approx 900$\,ld.
From their seven years long reverberation campaign
\cite{2000ApJ...533..631K} reported
Balmer line lags $\tau_{\rm cent}$ against the 5100\AA~ continuum of 
\hbox{H$\alpha$ $\sim 440$\,d,} \hbox{H$\beta$ $\sim 330$\,d,} \hbox{H$\gamma$ $\sim 265$d} 
(their Table~6, here converted to rest frame lags).

These lags are consistent with the lags for H$\beta$, H$\gamma\sim 260\,$d (rest frame) reported by
\cite{2019ApJ...876...49Z}; for compatibility we consider here the lag values
without detrending (listed in their Table~7).

\begin{figure}
  \centering
  \includegraphics[width=0.8\columnwidth]{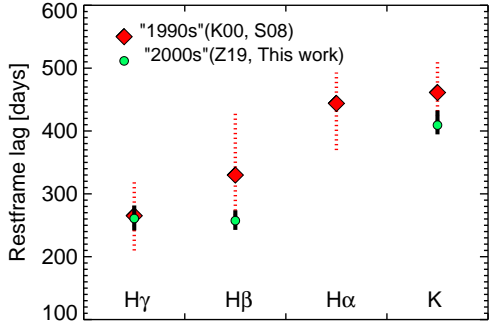} 
  \caption{Balmer and dust lags for 3C\,273 at different epochs: in the ``1990s'' 
    from K00 = \cite{2000ApJ...533..631K}, S08 = \cite{2008A&A...486..411S} and 
    in the  ``2010s'' from Z19 = \cite{2019ApJ...876...49Z} and this work. 
    The plotted lags are obtained with the ICCF method.
    \label{fig:lags_kaspi_zhang}
  }
\end{figure}

Figure~\ref{fig:lags_kaspi_zhang} presents the rest frame lags in days for H$\gamma$, H$\beta$, H$\alpha$ and NIR $K$-band for the monitor campaings in the ``1990s'' \citep{2000ApJ...533..631K,2008A&A...486..411S} and in the ``2010s'' (\cite{2019ApJ...876...49Z} and this work). In the ``1990s'' the dust lag is longer than the BLR lags, consistent with the unified scheme.

However, the difference between dust lag and BLR lags is small, in particular for H$\alpha$.
This may be explained -- at least partly -- in the bowl model by the foreshortening effect.
While the dust lag suffers from a strong foreshortening effect, 
the foreshortening of the BLR echo depends on how much above the equatorial plane 
the BLR clouds are located inside the bowl (Fig.~\ref{fig:bowl_K_lc}, left).
The similarity of the H$\alpha$ lag with the dust lag suggests that
H$\alpha$ emitting gas lies close to the dust
emitting bowl rim, in front of the rim as seen from the AD. 
We come back to that interesting possibility in Sect.~\ref{subsec:r_l_relation}.
Alternatively, one would have to take the much smaller BLR lags derived after a
detrending (of the optical continuum light curves),
e.g. $\tau$H$\beta$ $\sim 150\,$d \citep{2019ApJ...876...49Z}.
We note that the dust lags essentially remain unchanged, if a detrending is applied,
as we checked with several tests.
A detailed investigation of these issues will be presented in a forthcoming paper.


Finally we note a direct consequence of the dust torus geometry for cosmological applications.
For the nearby Sy-1 NGC\,4151 \cite{2014Natur.515..528H} calculated a dust-parallax distance,
based on dust RM data and interferometric size measurements.
Likewise the \cite{2019arXiv191000593G} tried that for three AGN (Mrk\,335, Mrk\,509 and NGC\,3783),
however with an extreme scatter.
If the bowl model is true, then the lags should be converted to real $R_x$, taking
also into account the inclination of the bowl.\footnote{For a bowl model at
fixed $R_x$ and $\theta$ range, the dust lag strongly increases
with inclination, e.g. for $i = 0^\circ$ and $i = 45^\circ$ the (simple geometric) TFs yield
$\tau_{45^\circ} \sim 2 \times \tau_{0^\circ}$,
because the TF is dominated by that side of the bowl, which is tilted away from the observer.
This questions a widely made assumption \citep{2014Natur.515..528H}:
``For reverberation mapping, however, inclination only
broadens or smooths the time lag signal symmetrically around the mean without a
significant shift in $\tau$.''
Nevertheless, NGC\,4151 lies at $i \sim 45^\circ$, so that in the  bowl model $R_x \sim \tau_{45^\circ}$ and
the derived parallax distance should be correct.}
Im principle,  parallax distances could be derived for 3C\,273 as well,
but we think that the uncertainty of the current data is
by far too large for allowing a reliable angular distance calculation.

\subsection{On the lag--luminosity relation}
\label{subsec:r_l_relation}

Figure \ref{fig:K_L} (top) shows the lag--luminosity diagram
for two different NIR dust RM data sets, one from our OCA campaigns and
one from the MAGNUM observations \citep{2014ApJ...788..159K,2019arXiv191008722M}, henceforth denoted with K14 and M19.


The analysed and published dust RM observations from OCA are on four sources
(PGC\,50427, WPVS\,48, 3C\,120, and 3C\,273).
All lags refer to $\tau_{\rm cent}$.
WPVS\,48 was observed during two independent campaigns in 2013 and 2014 
yielding -- within the errors -- the same lags \citep{2014A&A...561L...8P,catalinamaster}; here we take the average lag.
Our dust reverberation campaign of 
3C\,120 took place in 2014 -- 2015, one year after the factor 3 brightness outburst in 2013
which lasted until 2016 \citep{Ramolla2015,2018A&A...620A.137R}.
Within the short time span between the begin of the outburst and our reverberation campaign, 
the dust geometry might not have changed significantly and any large size changes are unlikely.
Therefore, 
we corrected the luminosity measured in 2014 -- 2015 down by factor 3 to match the luminosity before the outburst.
Table~\ref{tab:oca_dust_rm} lists the rest frame lags and luminosities used. A linear fit to the four sources (blue data points in Fig.~\ref{fig:K_L}, top)
yields a slope $\alpha = 0.33 \pm 0.01$ for the  lag--luminosity relation.
For comparison, the black dashed line marks a slope with $\alpha = 0.5$,
which is widely adopted \citep{Barvainis87,2014ApJ...788..159K,2014ApJ...784L..11Y,2019arXiv191008722M}.

Fig.~\ref{fig:K_L} shows also
the MAGNUM dust reverberation data from K14 and M19
as red squares and stars, respectively.
These lags were derived using the JAVELIN software \citep{2011ApJ...735...80Z};
while JAVELIN lags are basically similar to other CCF lags,
we do not know whether a bias could be present
and therefore we here consider the MAGNUM lags separately from the OCA lags.
The MAGNUM sample comprises 41 sources,
17 sources from K14 and 24 sources from M19,
making it the largest homogeneously obtained dust RM set.
All data are re-analysed by M19; we used the observed lags from their Table 3 column 3
(labeled  $\alpha_{OIR} = 1/3$, the optical -- NIR power-law index of the AD) 
and their Table 6, and
corrected the observed lags for time dilation $1/(1+z)$.
Strikingly, a linear fit to these MAGNUM data (all red points) yields a 
slope $\alpha = 0.34 \pm 0.03$. In Figure~\ref{fig:K_L} are also shown the 
residuals (data/fit); bottom: left for slope 0.5, right for slope 0.34.
Fitting all MAGNUM and OCA sources together yields a slope $\alpha = 0.339 \pm 0.024$.
A fit excluding the three sources with $log(L)> 45$~erg/s yields $\alpha = 0.338 \pm 0.030$.
Thus, the slope is not biased by a few luminous sources.\footnote{In their paper 
on the C\,IV $\lambda$1549 lag-luminosity relation, 
\citet{1991ApJ...370L..61K} noticed the exceptional position of 3C\,273 
with respect to a slope $\alpha = 0.5$ (see their Fig.~1). 
In that paper, 3C\,273 was the only source at the luminous end of the relation.
A simple check on the reliability of a relation is to remove the four extremes, 
each one at the top, bottom, left and right of the diagram. 
Consequently, to bring the position of this single source into agreement with $\alpha = 0.5$, they suggested
that the luminosity of 3C\,273 is an outlier and could be enhanced by beaming of continuum associated with the radio source. 
This possibility can largely be ruled out here, with the help of the 
two additional luminous radio-quiet sources in the sample of M19 (Fig.~\ref{fig:K_L}). 
} 

\begin{figure}
\centering
  \includegraphics[width=\columnwidth]{koshida_minezaki_dust_lum_relation.ps_page_1}

  \hspace{-1mm}\includegraphics[width=0.5\columnwidth]{koshida_minezaki_dust_lum_relation.ps_page_2}
  \hspace{-1mm}\includegraphics[width=0.5\columnwidth]{koshida_minezaki_dust_lum_relation.ps_page_3}

  \caption{$Top$: Dust lag versus $V-$band luminosity of AGN.
    Blue dots are data from OCA obtained by our group:
    PGC\,50427 \citep{2015A&A...576A..73P}, WPVS\,48 \citep{2014A&A...561L...8P},
    3C\,120 \citep{2018A&A...620A.137R}, and 3C\,273 in this work.
    The luminosity of 3C\,120 has been scaled by a factor 0.33,
    to account for the brightness outburst by a factor 3 in 2013 -- 2015.
    The red data points are from \citep{2014ApJ...788..159K} and \citep{2019arXiv191008722M},
    whereby we used the observed lags for power-law AD slope +1/3 and corrected for the time dilation.
    The black dashed line is the $\tau - L$ slope  with $\alpha = 0.5$ widely used.
    The blue and red lines are fits to the blue and red data points, respectively,
    both yielding a slope $\alpha = 0.34$ as labelled.
    $Bottom$: Residuals data / fitted line for $\alpha = 0.5$ (left) and $\alpha = 0.340$ (right).
      \label{fig:K_L}
  }
\end{figure}

\begin{figure}
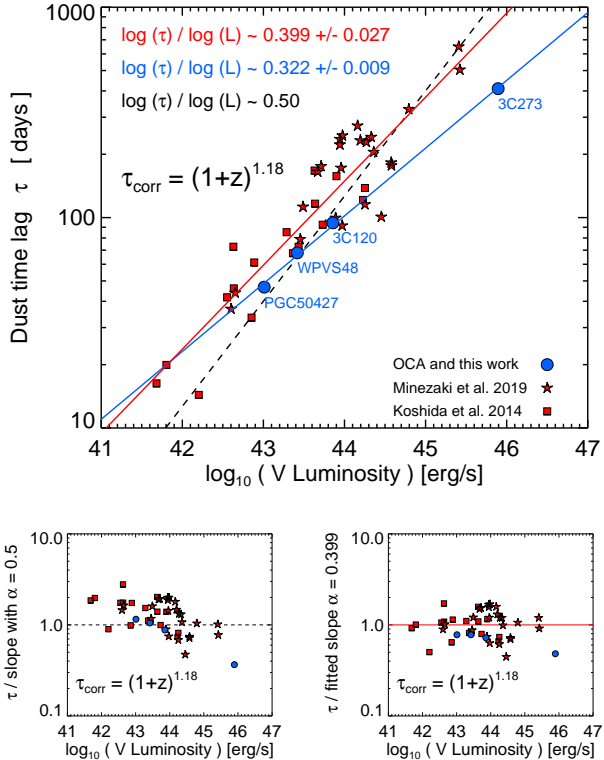

  \centering
  \includegraphics[width=\columnwidth]{koshida_minezaki_dust_lum_relation_with_tau_corr.ps_page_1}

  \hspace{-1mm}\includegraphics[width=0.5\columnwidth]{koshida_minezaki_dust_lum_relation_with_tau_corr.ps_page_2}
  \hspace{-1mm}\includegraphics[width=0.5\columnwidth]{koshida_minezaki_dust_lum_relation_with_tau_corr.ps_page_3}

  \caption{Same as Fig.~\ref{fig:K_L} but with rest wavelength correction of
    the lag $\tau$ by a factor (1+$z$)$^{\rm 1.18}$ from \cite{2019arXiv191008722M}.
    \label{fig:K_L_with_tau_corr}
  }
\end{figure}

When observing in a fixed NIR band, the rest-frame wavelength of the observed dust emission becomes shorter
at larger redshifts.
In an attempt to account for this,  \cite{2019arXiv191008722M}  derived a sophisticated
wavelength-dependent correction for the lags, by multiplying with a redshift term $\tau_{\rm corr} = (1+z)^{1.18}$.
These wavelength-dependent corrected rest frame lags are listed in their Table 3 column 6.
In the lag--luminosity diagram (Fig.~\ref{fig:K_L_with_tau_corr})
the lags become larger than without that correction (Fig.~\ref{fig:K_L}).
The correction shifts the luminous sources more upwards, because they are typically at higher redshift
(up to $z =  0.6$) than the low luminosity sources.
We applied the correction also to our OCA data, shown with blue colors in Fig.~\ref{fig:K_L_with_tau_corr}.
The fitted slope of 0.33 $\pm$ 0.01 remains about the same as without correction,
because all OCA sources are at small redshift ($z < 0.158$).
We fitted  the corresponding lag--luminosity relation for the different data sets, yielding slopes
of $0.39 \pm 0.045$ (for K14 only), $0.37 \pm 0.050$ (for M19 only),
$0.40 \pm 0.027$ (for the combined K14 and M19 data), and
$0.38 \pm 0.028$ (for the combined OCA and MAGNUM data).
These slopes are steeper than without the wavelength-dependent correction,
but significantly (at the $3 \sigma$ level) shallower than the slope 0.5.
At the high luminosity range ($\log{L} > 45$\,erg/s), 3C\,273 shows a
relatively small lag compared to the two other quasars
(PG\,0953$+$414, SDSS\,J0957$-$0023) but within the scatter
(see residual plot bottom right of Fig.~\ref{fig:K_L_with_tau_corr}).
\cite{2019arXiv191008722M} already noticed the exceptional position of 3C\,273 
based on the lags by \cite{2008A&A...486..411S} and tentatively attributed it to 
the radio loudness of 3C\,273. 
However, the two other quasars are radio quiet 
and a luminosity enhancement by the optical emission of a radio component 
does not explain the shallow slopes.

A matter of a debate is the wavelength dependence of the dust reverberation lag,
here considered versus the optical $UBVR$ bands, most commonly the $B-$ or $V-$band.
While the $J-$band is more sensitive to hotter dust than the $H-$ and $K-$bands, it appears reasonable to
expect a mix of hot dust temperatures for each spatial location \citep{2015OAP....28..175O}.   
There is no doubt that mid-infrared (MIR) lags are longer than NIR lags,
which led to the common interpretation that the cooler
dust emission arises from larger distance to the central heating source.
Several groups found a longer lag at $L$ ($\sim 3.6$\,$\mu$m) or $M$ ($\sim 4.8$\,$\mu$m)
compared to the $J-$ or $K-$band, e.g. \cite{Glass2004,2018rnls.confE..57F,Lyu2019}.
Based on sophisticated modeling of very sparse WISE observations \cite{Lyu2019}
report a lag ratio $K:L:M \sim 0.6 : 1 : 1.2$,
while \cite{2018rnls.confE..57F} find $J: K:L:M \sim 0.7 :  0.7 : 1.1 : 1.2$ for the Seyfert WPVS\,48
employing the combination of ground based $J,K$ and $Spitzer-IRAC1/2$ monitoring,
Notably, neither \cite{Glass2004} nor \cite{2018rnls.confE..57F} found significant differences in the lags
between rest frame 1 and 2\,$\mu$m;
typically any NIR lag differences are less than 5\% of the optical-NIR lag and not significant at the $3 \sigma$ level.
 The same holds for several other AGN with dust lags jointly determined
in $JK$. 
Therefore, we believe that the wavelength-dependent lag correction applied by M19 should be adopted with care.
In any case, both the OCA and MAGNUM data indicate a shallow slope between 0.33 and 0.4.
This questions the widely adopted lag--luminosity slope of 0.5. 



\begin{table}
  \centering
  \caption{OCA dust RM sample.
    \label{tab:oca_dust_rm}
  }
  \begin{tabular}{c c c c}
    \tableline
    Object   & z & $\tau_{\rm K, rest}$ [days] & $log(L_{V})$ [erg/s]  \\ \tableline
    PGC\,50427 & 0.024 & $46.2 \pm 2.6 $ & $43.0   \pm 0.12 $  \\
    WPVS\,48   & 0.037 & $68   \pm 5  $  & $43.40  \pm 0.10 $  \\
    3C\,120    & 0.033 & $95   \pm 6  $  & $43.84^*\pm 0.19 $  \\
    3C\,273    & 0.158 & $410  \pm 40  $ & $45.82  \pm 0.15 $  \\ \tableline
    \\
  \end{tabular}
  \\
  $^*$ brightness reduced by factor 3 to get the pre-outburst luminosity
\end{table}

The red data points show a large scatter around the fitted lag--luminosity relation
(red line in Fig.~\ref{fig:K_L}).
In the frame of the bowl model this could --  at least partly -- be understood as an effect of the inclination:
If the dust emission originates in general from the edge of the bowl rim and if
the AGN have different inclinations, then the TFs show a spread in shapes and asymmetries. For large inclination,
the TF becomes dominated by that side of the bowl, which is tilted away from the observer
(Figure~\ref{fig:bowl_K_lc},left).
This will shift the measured lags to larger values between optical and NIR light curves.
In parallel, for large inclination ($i > 20^\circ$) , one may expect that the observer's line-of-sight
to the central AD  crosses more absorbing material so that
the luminosity may be reduced, compared to inclination $i = 0^\circ$ (face-on view).
Together, in the lag--luminosity plot, the source will shift up and left from the actual relation,
inevitably leading to a scatter in the observed lag--luminosity data points.
To reduced the scatter, it would be desirable to obtain
via bowl modeling of the data a relation between $R_x$ and $L$, provided the data
are of sufficient quality.
We suggest that the scatter in such a size--luminosity relation may be reduced.
This is essential for cosmological applications like quasar distance estimates
\citep{1998SPIE.3352..120K, 1999OAP....12...99O, 2002ntto.conf..235Y,
  2014ApJ...784L..11Y, 2014ApJ...784L...4H,2019arXiv191008722M}.

 In search for possible explanations for a shallow R-L slope, 
  one possibibilty could be that luminosity-dependent internal extinction plays a role. 
  \cite{2004ApJ...616..147G} derived the reddening curve for AGN. From the examined data sets they also found that, 
  on average, low luminosity AGN ($logL_{\rm opt} \sim 44$ erg/s) are redder and 
  suffer from larger extinction ($A_V \sim 2.5\,$mag) than high luminosity AGN ($logL_{\rm opt} \sim 46$ erg/s), 
  see their Figure~5. 
  Then in the R-L relation an intrinsic slope $\alpha = 0.5$ could be 
  tilted to a shallower slope like the one we have observed here. 
  While this is an elegant straight forward explanation, it is worth to have a closer look. 
  If the extinction of Seyfert-1 nuclei is as large as $A_V \sim 2.5\,$mag and occurs in a screen between the BLR and the observer, 
  then one would expect to see its signature also in the Balmer line decrement.
  With a few individual exceptions, however, the flux ratio, $R_{\rm Balm}$, of the broad H$\beta$ and H$\alpha$ lines of 
  both quasars and Seyfert-1s lies in the typical range about $R_{\rm Balm} \sim 3$, 
  so that screen extinction of $A_V \sim 2.5\,$mag in Seyfert-1s may be unlikely, hence rejecting simply screen extinction.
  Then the case of mixed extinction remains, where the absorbing dust has to be mixed with the BLR emitting gas, 
  and the extinction affects essentially the background emission, which in our case is the continuum emission from the AD. 
  For mixed extinction -- likewise as in the presence of scattering material \citep{Krugel_2009} -- the amount of 
  extinction is mostly underestimated, 
  because the dust--gas mixture is considered only to small optical depths 
  ($\tau_{\rm opt} < 1$, here not to confuse with the lag $\tau$).
  Then this surface treatment of the Balmer line decrement mimicks a smaller than real extinction. 
  However, in case of such strong mixed extinction, the Balmer lags are expected to be similar to the hot dust lags, 
  contrary to the lags observed for Seyfert-AGN so far. 
  Another check of a high nuclear extinction in Seyfert-1s may be offered by the Flux Variation Gradients (like the ones shown in 
  Fig.~\ref{fig:fvg_bv}). They measure the continuum slope at wavelengths around the blue bump but so far do not show any luminosity 
  dependence \citep{1992MNRAS.257..659W}.
  Future studies are needed to clarify these issues. 


Finally we address two more possible lines for explaining the shallow lag--luminosity slope between 0.33 and 0.4.

%



1) Str\"omgren-behaviour:  We consider the case that the lag--luminosity relation implies
a relation between luminosity and the sublimation radius $R_{\rm sub} = f \cdot c \cdot \tau \propto L^{\alpha}$ where slope $\alpha = 1/3$ (and $c$ is the speed of light and $f$ is a scaling factor).
The relation with slope $\alpha = 1/3$ is strikingly reminiscent to the
well known size--luminosity relation for H\,II regions, where
$R_{\rm H\,II} \propto L^{1/3}$ \citep{Stroemgren_1939ApJ....89..526S,Mc_Cullough_2000}.
The analogy between $R_{\rm sub}$ and $R_{\rm H\,II}$ is as follows:

For H\,II regions the  Str\"omgren radius $R_{\rm H\,II}$ describes up to which distance
from the ionising star the radiation field is strong enough to ionize, e.g. the hydrogen atoms.
Beyond $R_{\rm H\,II}$ the radiation field is too weak so that the atoms ``survive'' unaffected.
The reason for the slope $\alpha = 1/3$ for the Str\"omgren relation is
that interjacent material inside $R_{\rm H\,II}$  absorbs the radiation from the ionising star.

For the dust in AGN we
deal with the sublimation radius $R_{\rm sub}$.
Inside of $R_{\rm sub}$ the AGN radiation field is sufficiently strong,
so that the dust grains evaporate (in analogy to become ionized). Outside of $R_{\rm sub}$
the radiation field is too weak so that the dust grains can survive.
So far, the assumption was made that the (dust heating UV) photons of the AD
travel all the way until $R_{\rm sub}$ without being absorbed by interjacent material.
This led to the widely adopted relation $R_{\rm sub} \propto L^{1/2}$.
However, if sufficient absorbing material lies between the AD and the dust grains,
then $R_{\rm sub}$ becomes smaller, and this will lead to a shallower slope $\alpha < 1/2$.
Then the  slope $0.34 < \alpha < 0.4$ found in Figs.~\ref{fig:K_L} and \ref{fig:K_L_with_tau_corr}
implies that there is in fact plenty of absorbing material between the AD and the dust grains.
This means for the bowl model considered here,
that even the line between the AD and the bowl edge at $\theta \sim 45^\circ$ crosses a
significant amount of absorbing material.
The large H$\alpha$ lag / dust lag mentioned in Sect.~\ref{sec:dust_geometry} may hint to such material.

2) Geometric effects of a bowl mirror:
We assume a bowl model with fixed half opening angle 45$^\circ$ irrespective of the AD luminosity.
Clearly, both the bowl rim and material inside the bowl act in the net effect like a
reflecting mirror for the photons from the AD.
The photons are scattered by electrons and dust grains.
In addition, reprocessed continuum emission may play a role;
for instance \cite{2019NatAs...3..251C} found evidence of a non-disk optical continuum
emission around AGN, which likely comes from the inner wall of the BLR. 
Thus, the observer sees -- in addition to the AD flux $F_{\rm AD}$ -- also a contribution $F_{\rm bowl}$
from scattered or reprocessed photons.
This leads to an amplification of the original AD brightness.
The relative amplification $Ampl = F_{\rm bowl} / F_{\rm AD}$ may be small (a few percent) but it is worth to consider how far it is luminosity dependent.
We make the assumption that the bowl opening angle and the geometric covering angle
for intercepting AD photons is luminosity independent.
Then the effect of the bowl rim on $Ampl$ might be scale-invariant.
However, the volume inside the bowl increases proportional to $R_x^3$.
If the density of scattering or reprocessing particles inside the bowl is
independent of the AD luminosity,
then one may expect that $F_{\rm bowl}  \propto R_x^3$.
This yields $Ampl \propto R_x^3$.
Assuming for simplicity $R_x \propto L^{1/2}$ we get $Ampl \propto L^{3/2}$.
In other words, in the net effect the actual luminosity of the AD may be
overestimated by a factor which scales with $L^{3/2}$.
Then in the lag--luminosity relation the data points
will be shifted to
large $L$ values, so that
the resulting slope becomes shallower than $\alpha = 0.5$.

A detailed quantitative consideration of the potential
Str\"omgren-like behaviour of AGN and the geometric effects of a
bowl mirror will be presented in a future paper.



\section{Summary and conclusions}\label{sec:conclusions}

We performed  a 5 years reverberation mapping campaign of 3C\,273 in the optical
($BVrz$) and NIR ($JHK$) bands at
the Bochum University Observatory near Cerro Armazones (OCA).
The optical light curves were supplemented by longer and denser sampled $V-$band light
curves from \citet{2019ApJ...876...49Z}. The results are:

\begin{enumerate}
\item
  To obtain the pure dust light curves, the contribution of  host galaxy and
  accretion disk to the NIR bands had to be removed.
  The resulting dust light curves show consistently correlated variations in all three NIR bands,
  giving confidence that the procedure worked well to remove the contribution of host galaxy and accretion disk.

\item
  For all three filter pairs ($J/H$, $J/K$, $H/K$) the color temperatures change by
  about 5\% (i.e. a factor 1.05) between the bright and faint states.
  On the other hand, the amplitudes of the dust light curves increase with decreasing
  wavelength from 0.2 at $K$ to 0.4 in $J$.
  Because the filters measure the dust emission on the Wien tail of the Planck function,
  the brightness changes are expected to become larger at shorter wavelengths;
  they may exceed the amplitude of the triggering signal light curve.
  This altogether consistently indicates that the variations of the dust emission are likely
  due to (mean) temperature changes of the dust grains in the order of 5\%.

\item
  We derived the dust covering factor $CF$ from the optical/UV and NIR luminosities,
  yielding small values $CF\sim$ 8\%, consistent with the results $CF\sim$ 7\%
  for other type-1 AGN by \cite{2011MNRAS.414..218L}.

\item
 We determined the time lag $\tau$ of the dust light curves against the $V$-light curve trough different CCF methods and found an average time lag of $\tau_{\rm K, rest} \sim 410$\,d. Some correlation methods reveal an interesting asymmetry, which is consistent with the transfer function of a tilted dust geometry.

\item
 We re-analysed the data of \cite{2008A&A...486..411S} and compared the dust time lag and CCF asymmetries 
during our observing campaign and theirs.

\item
  The average time lag of $\tau_{\rm rest} \sim 410$\,d is a factor of $\sim 2$ smaller than expected from the interferometric ring radius of $\sim 900$\,ld.
  Interferometry measures the projected size as seen by the observer, while RM measures the 3-dimensional light travel time difference in the system.
  We suggest that the difference between interferometric size and RM lags
  can be explained by 3D geometrical effects, in particular the foreshortening
  effect from which the reverberation data suffer.

\item
  To bring the observational findings into a consistent picture, we
  considered a bowl shaped torus geometry as proposed by \cite{2012MNRAS.426.3086G},
  where the dust emission originates from the edge of the bowl rim with a small covering angle
  $40^\circ < \theta < 45^\circ$, as justified by the small $CF$.
    We used an inclination angle of $12^{\circ}$ indicated from radio jet
  studies \citep{2001Sci...294..128L, 2006A&A...446...71S, 2017ApJ...846...98J}.
  For such a model with an equatorial size  $R_x \sim 900$\,ld 
  the (simple geometric) transfer function (TF) is double-horned and yields an 
  asymmetric cross correlation similar to that found from the data.
  We have convolved different TFs for a bowl geometry
  with the host-subtracted $V-$band light curve 
  and showed the corresponding echo light curves.
  For an equatorial size  $R_x \sim 900\pm200$\,ld, 
  the echo light curve delays are
  in agreement with the delays found in our data, and the modeled cross correlations show an asymmetry similar to that observed in the CCF.
  The main conclusion from our study is that the hot dust emission seen in the NIR originates from a tilted ($i = 12^\circ$) thin ring which lies above the equatorial plane.


\item
  The relation between dust lag and optical luminosity shows a large scatter.
  To reduce the scatter for future cosmological applications,
  it may be desirable to obtain $R_x$  via modeling
  of the data (provided they are of sufficient quality)
  and check for a relation between $R_x$ and $L$.

\item
  We find for the lag--luminosity relation a rather shallow slope 
  between 0.33  and 0.4.
  This rejects the widely adopted slope 0.5 at the 3-sigma level.
  We envisage three possible explanations for the shallow slope:

  1) 
  \citet{2004ApJ...616..147G} found that the internal extinction in AGN increases with decreasing AGN luminosity 
  by $A_{\rm V} \sim 2.5$\,mag between quasars and Seyferts.
  If the internal extinction in AGN is in fact so high and shows such a strong luminosity dependence, 
  then the slope may be tilted from 0.5 to about 0.33.

  2) AGN have a similar behaviour as H\,II regions where the size of the ionized region $R_{H\,II} \propto L^{1/3}$.
  If a substantial amount of absorbing material lies between the AD and the dust grains,
  this allows for a shortened dust sublimation radius. If the material density is $L$-invariant,
  the relative shortening increases with the path length between AD and dust.
  The path length (bowl size) depends on $L$.
  Then, in the lag--luminosity diagram, the relative reduction of the lag increases with $L$.

  3) The observer measures an AD luminosity $L$ which is magnified by scattered and reprocessed
  radiation from material in the bowl,
  and the relative contribution of this magnification of $L$ increases with the volume of the bowl
  and therefore also with $L$.
  Then, in the lag--luminosity diagram, the relative overestimation of $L$ increases with $L$.

\end{enumerate}
While these findings apply to the single case of the luminous quasar 3C\,273, future detailed studies
on a larger quasar sample should be envisaged to corroborate the conclusions.

\acknowledgments
The project was supported by funds from the Akademie der Wissenschaften Nordrhein-Westfalen and Deutsche Forschungsgemeinschaft HA3555/12 and HA3555/14. The observations benefitted from the care of the guardians Hector Labra, Gerardo Pino, Roberto Mu\~{n}oz, and Francisco Arraya. We warmly thank Jian-Min Wang and Zhi-Xiang Zhang for sending us their light curves of 3C\,273, and the referee Martin Gaskell for his detailed constructive report.





\bibliography{Literature_all}{}
\bibliographystyle{aasjournal}

\clearpage
\appendix
\section{Different Cross Correlation Functions}\label{sec:appendix}
\begin{figure*}[h]
\centering
\includegraphics[width=16cm]{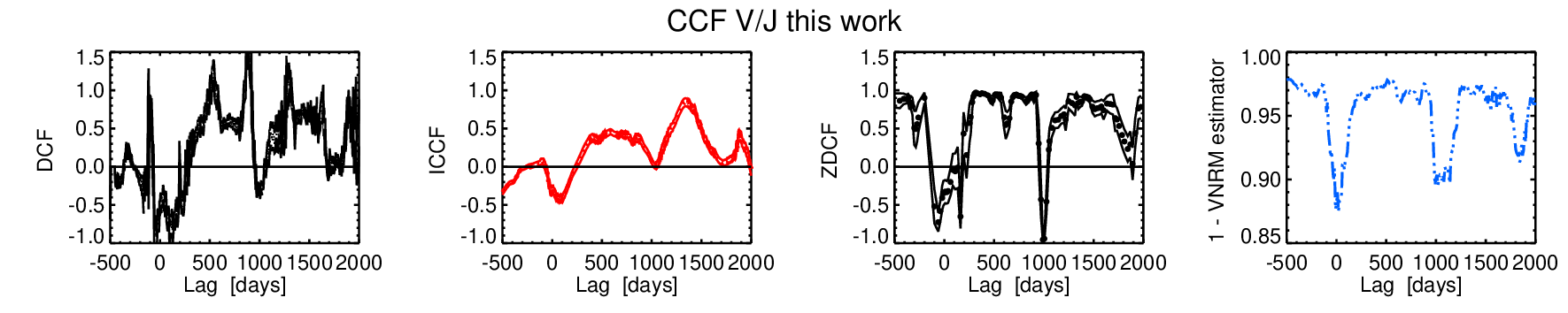}\\
\includegraphics[width=16cm]{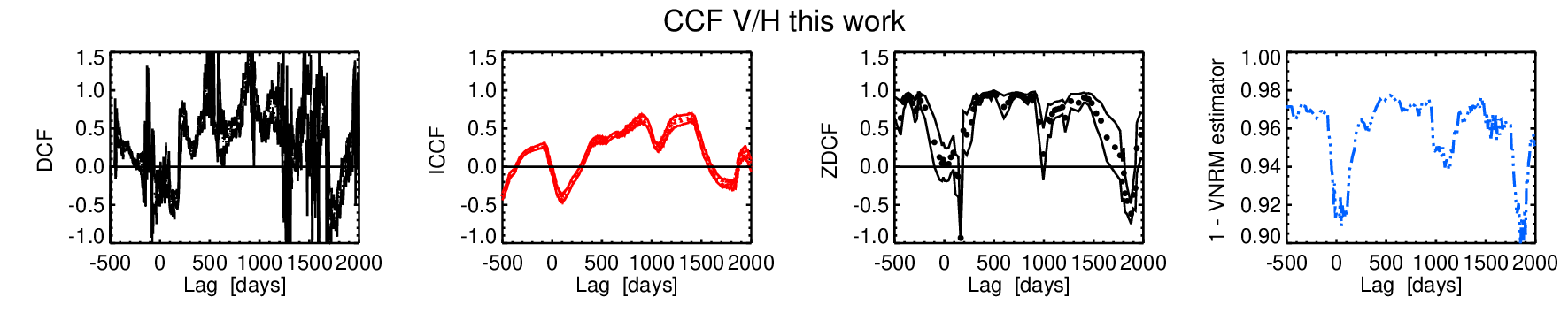}\\
\includegraphics[width=16cm]{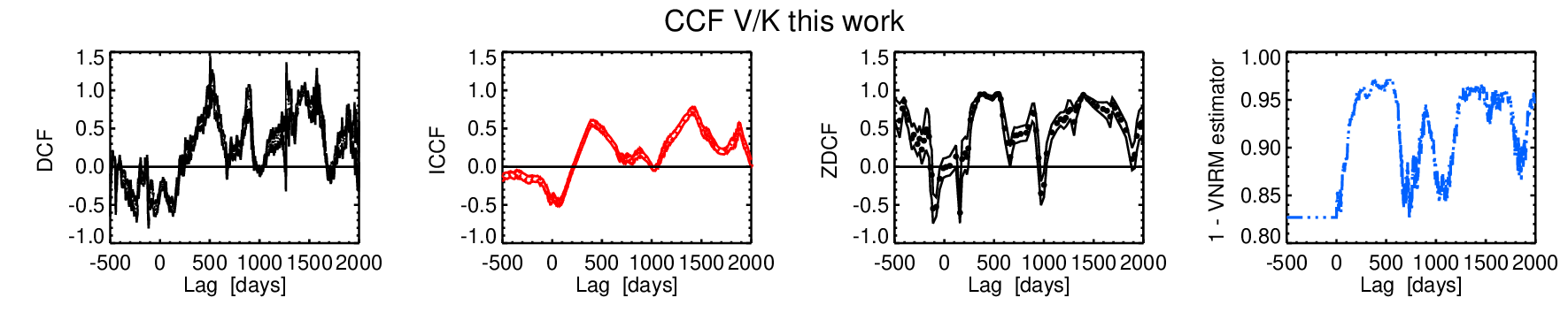}\\
\includegraphics[width=16cm]{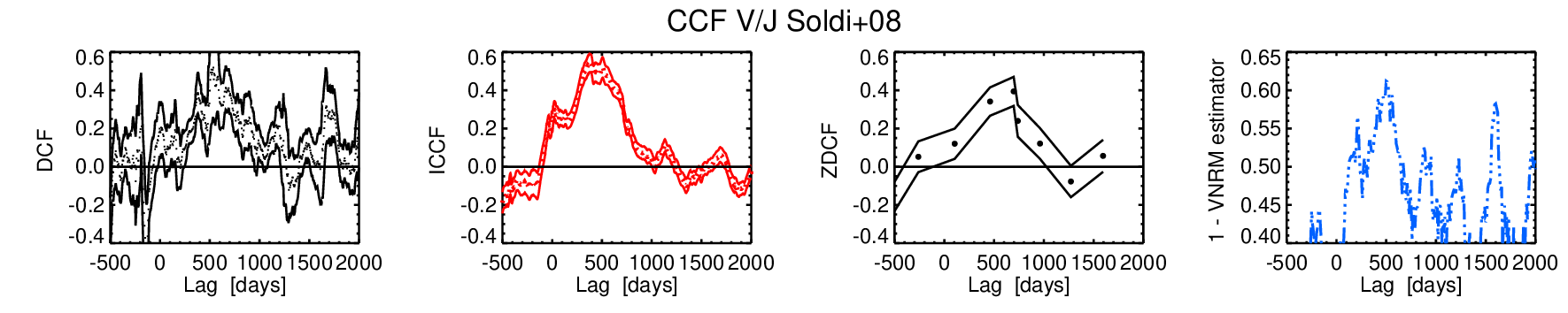}\\
\includegraphics[width=16cm]{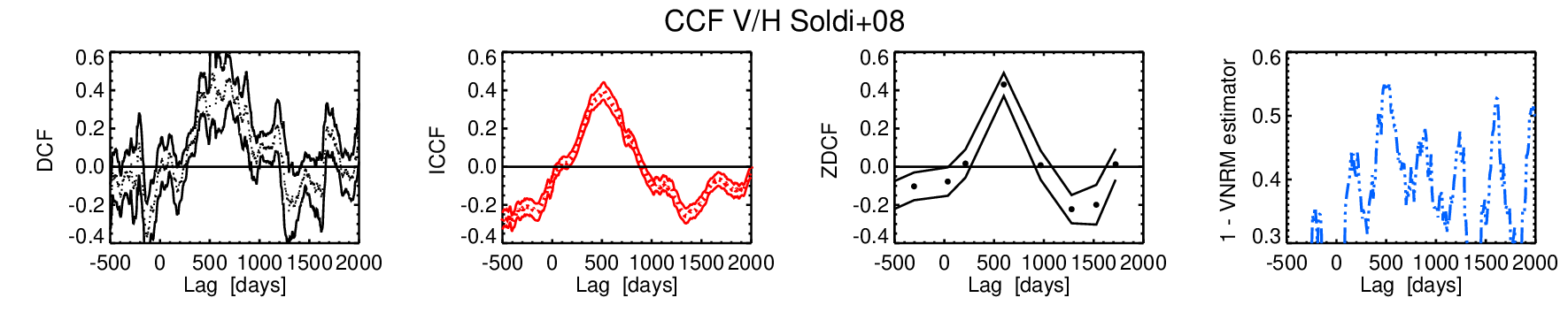}\\
\includegraphics[width=16cm]{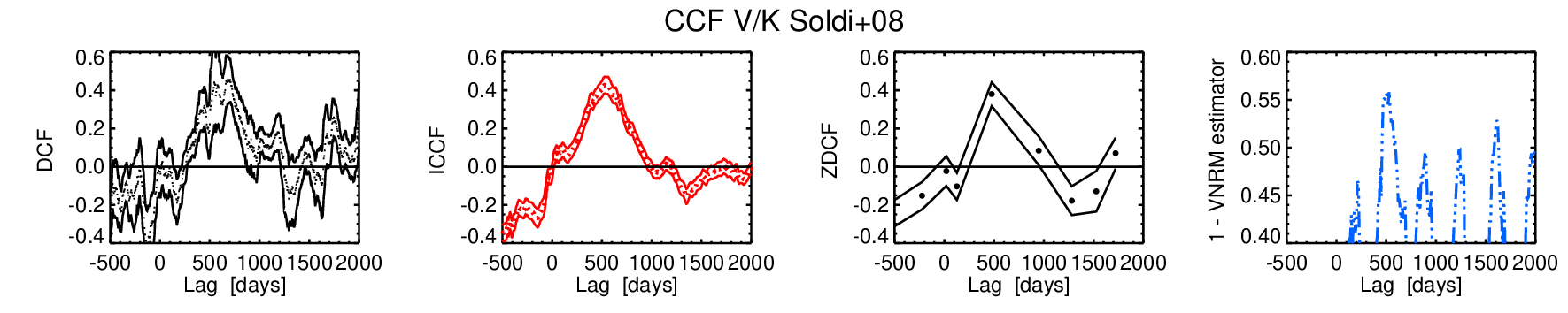}\\
\caption{DCF, ICCF, ZDCF and VNRM between $V$ and $JHK$ light curves for this work and \cite{2008A&A...486..411S}. }
\label{fig:ccf_appendix_soldi}
\label{fig:ccf_appendix_oca}
\end{figure*}

\end{document}